\documentclass[nologo,11pt,a4paper]{ETHpaper}
\pdfoutput=0

\usepackage{graphicx, epsfig, amsmath, amssymb}
\usepackage[square,numbers,sort&compress]{natbib}

\begin{document}

\newcommand{\mean}[1]{\left\langle #1 \right\rangle}
\newcommand{\abs}[1]{\left| #1 \right|} \newcommand{\la}{\langle}

\newcommand{\sect}[1]{Sect. \ref{#1}}
\newcommand{\eqn}[1]{eq. (\ref{#1})}
\newcommand{\eqs}[2]{eqs. (\ref{#1}), (\ref{#2})}
\newcommand{\pic}[1]{Fig. \ref{#1}} \newcommand{\name}[1]{{\rm #1}}

\renewcommand{\textfraction}{0.01} 
\renewcommand{\topfraction}{0.99}
\renewcommand{\bottomfraction}{0.99}
\renewcommand{\floatpagefraction}{0.99}

\renewcommand{\thefootnote}{ \fnsymbol{footnote} }

\title{Optimal migration promotes the outbreak of cooperation in
  heterogeneous populations}

\titlealternative{Optimal migration promotes the outbreak of cooperation
  in heterogeneous populations}

\author{Frank Schweitzer$^{\star}$, Laxmidhar Behera$^{\dagger}$}

\authoralternative{F. Schweitzer, L. Behera}

\address{$^{\star}$ Chair of Systems Design, ETH Zurich,\\ Kreuzplatz 5,
  8032 Zurich, Switzerland, \texttt{fschweitzer@ethz.ch}
  \\
  $^{\dagger}$ Department of Electrical Engineering, Indian Institute of
  Technology, Kanpur, 208 016, India, \texttt{lbehera@iitk.ac.in} }

\reference{\emph{ACS - Advances in Complex Systems}, vol. \textbf{15},
  Suppl. No. 1 (2012) 1250059 (27 pages)}

\www{\url{http://www.sg.ethz.ch}}

\makeframing
\maketitle

\renewcommand{\thefootnote}{\fnsymbol{footnote}} 

\begin{abstract}
  We consider a population of agents that are heterogeneous with respect
  to (i) their strategy when interacting $n_{g}$ times with other agents
  in an iterated prisoners dilemma game, (ii) their spatial location on
  $K$ different islands. After each generation, agents adopt strategies
  proportional to their average payoff received. Assuming a mix of two
  cooperating and two defecting strategies, we first investigate for
  isolated islands the conditions for an exclusive domination of each of
  these strategies and their possible coexistence. This allows to define
  a threshold frequency for cooperation that, dependent on $n_{g}$ and
  the initial mix of strategies, describes the outbreak of cooperation in
  the absense of migration. We then allow migration of a fixed fraction
  of the population after each generation. Assuming a worst case scenario
  where all islands are occupied by defecting strategies, whereas only
  one island is occupied by cooperators at the threshold frequency, we
  determine the optimal migration rate that allows the outbreak of
  cooperation on \emph{all} islands. We further find that the threshold
  frequency divided by the number of islands, i.e. the relative effort
  for invading defecting islands with cooperators decreases with the
  number of islands.  We also show that there is only a small bandwidth
  of migration rates, to allow the outbreak of cooperation. Larger
  migration rates destroy cooperation.

\emph{keywords:} {migration, cooperation, iterated prisoners dilemma}

\end{abstract}

\section{Introduction}

\emph{Human Migration}, i.e. the movement of large numbers of people out
of, or into specific geographical areas, is seen as one of the biggest
challenges that face the human societies in the 21st century. 
On one hand, part of the human population has reasons to \emph{emigrate}
into countries which provide a ``better'' life -- on the other hand,
industrialized countries cannot sustain their current situation without
the \emph{immigration} of people. The real problem arises because the
``demand'' and the ``supply'' side cannot be matched. Industrialized
countries fear that immigrants do not contribute to their further
economic growth but, on the contrary, deplete their wealth by taking
advantage of a social security, health, and educational system which they
did not contribute to.

If we move this problem on the more abstract level of a game-theoretical
model, we can distinguish between two types of agents: those
\emph{cooperating}, i.e. being able to integrate in a society and to
contribute to a common good, namely economic growth, and those
\emph{defecting}, i.e. without the ability to socially integrate and thus
depleting a common good at the cost of the cooperating agents. Certainly,
based on their past experience, agents can adapt, i.e. they can change
their strategy from defection to cooperation and vice versa dependent on
the payoff they receive in a given environment. The question for an
industrialized country would be then to define an optimal immigration
rate that (a) does not destroy the common good, and (b) allows agents to
adapt to the assumed cooperative environment within one or two
generations, even if they may have not immigrated with a cooperative
strategy.

The problems of cooperation and defection and the payoff-dependent
adoption of strategies have been discussed in the framework of the
Prisoner's dilemma (PD) and the iterated PD (IPD) game (see
Sect. \ref{2}).  With our paper, we add to this framework the ability to
migrate between different countries (``islands''). Our aim is to reveal
optimal conditions for the migration of agents such that cooperating
strategies can take over even on those islands where they were initially
not present.

We note that migration was previously studied in a game-theoretical
context by different authors \cite{helbing2008migration,
  jiang2010role}. Our work differs from these attemts in various
respects. First of all, we do not assume that migration is based on the
anticipated success \cite{helbing2008migration, helbing2009outbreak} --
this shifts the conceptual problem of the ``outbreak of cooperation'' to
proposing rules with non-local information such that two cooperators meet
at the same place, from which cooperating clusters can grow. We also do
not make migration dependent on local variables such as the number of
defectors in the neighborhood \cite{jiang2010role} or random choices of
``empty'' places \cite{helbing2008migration, jiang2010role}. In fact,
human migration is rarely targeted at less densely crowded places, it is
rather the opposite. Further, we do not assume one-shot games such as the
PD, but instead consider the IPD in which the number $n_{g}$ of repeated
interaction as well as the mix of up to 8 different strategies plays a
crucial role.

Eventually, we do not use an agent-based model in which update and
migration rules are freely defined, to study their impact on computer
simulations on a lattice. Our approach proposes a population based model
in which subpopulations are defined with respect to (a) their interaction
strategy, and (b) their spatial location. The consideration of separated
``islands'' allows a coarse-grained implementation of spatial structures
which is in between a lattice or network approach and a meanfield
description.  It is known that spatial structures have an impact on the
outbreak of cooperation \cite{roca2009effect, lozano2008mesoscopic,
  schweitzer02:_evolut}, but their influence varies with other degrees of
freedom, such as update rules, synchronization, interaction topology,
payoff matrix.

Therefore, in this paper we adopt mostly standard assumptions about the
interaction type (IPD with $n_{g}$ encounters) and interaction topology
(panmictic subpopulations), strategy adoption (replication proportional
to payoff), and migration (fixed fraction of the population). To
understand the basic dynamics, we first investigate ``isolated'' islands
(no migration) to find out about the conditions for the ``outbreak of
cooperation'' without external influence. This ``outbreak'' is defined as
the critical point (strategy mix, number of encounters $n_{g}$) beyond
which a whole island is going to be occupied by cooperating agents, if
agents adopt strategies proportional to their average payoff. Then, we
add migration between islands to this dynamics to find out under which
conditions the outbreak of cooperation can be enhanced. It is important
to note that migration does not distinguish between strategies or
islands, i.e. there are no better suited strategies for immigration, or
bad places with high emigration rates -- which we consider as artificial
assumptions to explain the ``outbreak of cooperation''.

To determine the robustness of our findings, we always consider worst
case scenarios, i.e. initial settings in which most islands have either
an entirely defective subpopulation, or at least a defective majority. We
further control for important parameters such as the pool of available
strategies $s$, the number of interactions $n_{g}$ or the number of
islands $K$, for which critical conditions are derived. Our finding that
migration is indeed able to boost the outbreak of cooperation is
remarkable both because it is based on minimal assumptions about the
dynamics and because it follows from a quite systematic investigation of
the underlying conditions.

\section{Rules of the game}
\label{2}

\subsection{Strategic interaction}
\label{sec:strat}

\begin{figure}[htbp]
  \begin{center}
    \includegraphics[width=6.0cm]{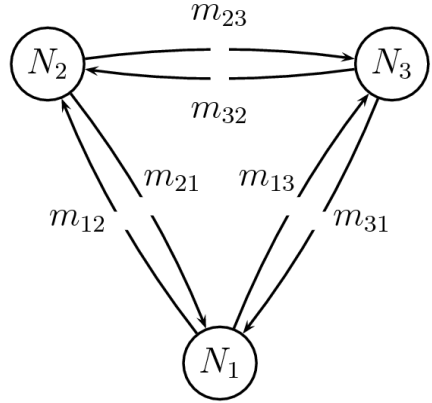}
  \end{center}
  \caption{Subpopulations $N_k$ (with different strategies) are spatially
    distributed on $k$ islands from where they can migrate at rates
    $m_{uv}$.}
  \label{fig:lattice}
\end{figure}

Let us investigate a population of $N$ agents divided into subpopulations
on $K$ different islands which imply a coarse-grained spatial structure,
i.e. $N=\sum^{K}_{k}N_{k}$.  Agents at island $k$ are assumed to interact
with the $N_{k}-1$ other agents on their island (\emph{panmictic
  population}).  But, in general we assume that migration between the
islands is possible, with the respective migration rates given by
$m_{uv}$. These define the fraction of the agent population at island $u$
that migrates to island $v$ in a given time interval. Figure
\ref{fig:lattice} shows the case of 3 different locations.

Our model basically considers two different time scales for interaction
and migration.  We define a \emph{generation} $G$ to be the time in which
each agent has interacted with all other $N_{k}-1$ agents a given number
of times, denotes as $n_{g}$. Thus the total number of interactions
during each generation in the panmictic population is roughly
$N_{k}^{2}/2 \times n_{g}$.  At any given time, agent $i$ can interact
only with one other agent in a ``two-person game''. But during one
generation $G$, it interacts with all other $N_{k}-1$ agents $n_{g}$
times (repeated two-person game).

For the interaction, we adopt the well known \emph{iterative prisoner's
  dilemma} (IPD) from game theory. At each encounter between agents $i$
and $j$, both have two actions to choose from, either to
\emph{collaborate} ($C$) or to \emph{defect} ($D$), \emph{without
  knowing} the action chosen by their counterpart. The outcome of that
interaction is described in the following payoff matrix:
\begin{equation}
  \label{2payoff}
  \begin{array}[c]{ccc}
    & \textbf{C} & \textbf{D}\\
    \textbf{C} & \;(R,R)\; & \;(S,T)\; \\
    \textbf{D} & \;(T,S)\; & \;(P,P)\;
  \end{array}
\end{equation}
If both agents have chosen to collaborate, then they both receive the
payoff $R$. If one of the agents chose to defect and the other one to
cooperate, then the defecting agent receives the payoff $T$ and the
cooperating the payoff $S$. If both defect, they both receive the payoff
$P$. In the special class of \emph{Prisoner's Dilemma} (PD) games, the
payoffs have to fullfil the following two inequations:
\begin{equation}
  \label{pd-ineq1}
  T > R > P > S\;; \quad 2\,R > S+T
\end{equation}
which are met by the standard values $T=5$, $R=3$, $P=1$, $S=0$. This
means in a cooperating environment, a defector will get the highest
payoff. 

For $n_{g}=1$ (``one-shot'' game) choosing action $D$ is unbeatable,
because it rewards the higher payoff for agent $i$ no matter if the
opponent chooses $C$ or $D$. At the same time, the payoff for \emph{both
  agents $i$ and $j$} is maximized when both cooperate.  A simple
analysis shows that defection is a so-called \emph{evolutionary stable
  strategy} (ESS) in a one-shot PD.  If the number of cooperators and
defectors in the population is given by $N^{c}$ and $N^{d}=N-N^{c}$
respectively, then the expected average payoff for cooperators will be
$a^{c} = (R \times N^{c} + S \times N^{d})/N$. Similarly the expected
average payoff for defectors will be $a^{d}= (T\times N^{c} + P \times
N^{d})/N$.  Since $T>R$ and $P>S$, $a^{d}$ is always larger than $a^{c}$
for a given number $N^{c}$, and pure defection would be optimal in a
one-shot game. Even one defector is sufficient to invade the complete
population of $N-1$ cooperators.

But in a \emph{consecutive} game with short \emph{memory}, both agents,
by simply choosing $D$, would end up earning less than they would earn by
cooperating. Thus, the number of games $n_{g}$ two agents play together
becomes important.  This makes sense only if the agents can remember the
previous choices of their opponents, i.e. if they have a memory of
$n_{m}\leq n_{g}-1$ steps. Then, for the \emph{iterated Prisoner's
  Dilemma} (IPD), they are able to develop different \emph{strategies}
based on their past experiences with their opponents.  We note that the
IPD game was studied both in the context of a panmictic population
(\citep{
Rapoport:96, Michael:96, Fogel:95, Vainstein:02,
  Axelrod:81, 
  Doebeli:97}) and assuming a spatial population structure
\citep{Cohen:99, Szabo:98, szabo00:_spatial}.

Usually, in an IPD only a \emph{one-step} memory is taken into
accout. Based on the known previous choice of its opponent, either $C$ or
$D$, agent $i$ has then the choice between \emph{eight} different
strategies. Following a notation introduced in \citep{Nowak:92}, these
strategies are coded in a 3-bit binary string
$\left[I^{o}|I^{c}\,I^{d}\right]$ which always refers to
\emph{cooperation}.  The first bit represents the \emph{initial} choice
of agent $i$: it is $1$ if agent $i$ has \emph{cooperated}, and $0$ if it
has defected initially.  The two other values refer always to the
previous choice of agent $j$.  $I^{c}$ is set to $1$ if agent $i$ chooses
to cooperate given that agent $j$ has cooperated before and $0$
otherwise. $I^{d}$ is similarily set to $1$ if agent $i$ chooses to
cooperate given that agent $j$ has defected before and $0$ otherwise.

Both $I^{c}$ and $I^{d}$ can be also interpreted as \emph{probabilities}
to choose the respective action given the knowledge of the previous
choice of the opponent, i.e. $\{0\leq I_{p},I^{d}\leq 1\}$ in the
stochastic case. In the \emph{deterministic} case considered in this
paper, $I^{d}$ and $I^{c}$ are either 0 or 1. Out of the eight possible
strategies, to keep our results in a tractable manner, we consider the
following four strategies $z^{(s)}$ ($s=1,2,3,4$):
\begin{equation}
  \label{strat}
  \begin{array}[c]{rrrcl}
    z^{(1)}:&& [1|10] & & \mathrm{TFT} \\
    z^{(2)}: &\quad & [0|00] & \quad & \mathrm{ALL-D} \\
    z^{(3)}: & & [1|11] &  & \mathrm{ALL-C} \\
    z^{(4)}: && [0|01] & & \mathrm{A-TFT} 
  \end{array}
\end{equation}
These four strategies were also considered originaly in the famous paper
by Axelrod \& Hamilton \citep{Axelrod:81}. Hence, we will extend these
investigations by later adding migration to the game. We note that all
eight strategies have been discussed in \citep{Schweitzer2005}. In this
paper we have dropped those variants that differ only in the first move
but then continue as described in the following.  The strategies ALL-D
and ALL-C are obvious: for $z^{(2)}$ agent $i$ always chooses to defect
regardless of the previous choices of agent $j$, $z^{(3)}$ follows
likewise. These two strategies represent the limit case of \emph{no
  memory}, i.e. agent $i$ is simply ``fixed'' as a cooperator or a
defector.

The most interesting strategy $z^{(1)}$, known as ``tit for tat'' (TFT),
means that agent $i$ initially cooperates and continues to do so, given
that agent $j$ was also cooperative in the previous move. However if
agent $j$ was defective in the previous move, agent $i$ chooses to be
defective, too. This strategy was shown to be the most successful one in
iterated Prisoners Dilemmas \citep{Axelrod:81}.  If initially two
cooperative agents meet, they stay cooperative (forever, in a
deterministic game), this way maximizing their average payoff. This
encourages other agents to adopt TFT. Hence, the ``outbreak of
cooperation'' in a community of $N$ agents occurs as the amplification of
a cooperative initial fluctuation. 

The fourth possible strategy $z^{(4)}$ is just the negation of the most
successful one, therefore named Anti-TFT. It means that agent $i$ starts
as a defector and continues to be so, as long as it meets with
cooperative agents $j$, this way receiving the highest possible payoff
$T$. But, if it meets an agent that has previously defected, it changes
its ``behavior'' to cooperation. This makes some sense, since the loss in
payoff is not that much -- $P$ would be just slightly higher than $S$.
On the other hand, if the colloborative agent meets again with another
cooperative agent, the payoff would be $R$, which is a much higher gain.
So, this strategy, while not making sense in the first place, may end up
with a higher payoff in those cases where defecting and cooperating
strategies are equally present. We will later show that A-TFT in the long
run benefits agents playing ALL-D and therefore will create a more
difficult environment for the invasion of cooperation. I.e. considering
all four strategies captures a worst-case scenario for the outbreak of
cooperation which we deem interesting to study (instead of giving
rational arguments in favor of A-TFT).

\subsection{Dynamics of replication and migration}
\label{sec:rep}

In this paper, we are interested in how the frequencies of the different
strategies would evolve in the agent community. The fraction of agents
choosing strategy $z^{(s)}$ on island $k$ is defined as
$f_{k}^{(s)}(t)=N_{k}^{(s)}(t)/N_{k}$, where the different strategies are
given by \eqn{strat}.  Similarly, the \emph{total fraction} of agents
playing strategy $z^{(s)}$ in the \emph{whole} system, is given as
$f^{(s)}(t)= N^{(s)}(t)/N$. For the dynamics we may assume that during
each generation $G$ every agent behaves according to a fixed strategy
picked up from the pool of possible strategies. This strategy can be
changed only after one generation is completed. The ``update'' rule for
the strategy is simply given by the success of different strategies
during the past generation. As the evaluation criteria, we choose the
average payoff $a_{k}^{(s)}$, each strategy has received during the last
generation on island $k$. This will be compared with the total average
payoff $\bar{a}_{k}$ that gives an estimate of the overall dynamics on
that island:
\begin{equation}
  \label{a-bar}
  \bar{a}_{k}(t)=
  \sum_{s=1}^{4} a_{k}^{(s)}(t) f_{k}^{(s)}(t)
  \;; \quad \sum_{s=1}^{4}
  f_{k}^{(s)}(t) =1 
\end{equation}
For the evolution of the frequencies of the different strategies, the
following dynamics is postulated:
\begin{equation}
  \label{a-dyn}
  f_{k}^{(s)}(G+1)=\frac{a_{k}^{(s)}(G)}{\bar{a}_{k}(G)}\;f_{k}^{(s)}(G)
\end{equation}
It means that in the next generation $G+1$ the share of agents choosing a
particular strategy $z^{(s)}$ has grown/shrunk -- i.e., strategy $z^{(s)}$ has
replicated -- according to the \emph{relative} performance of this
strategy on island $k$ during the previous generation $G$. In population
dynamics, this is known as \emph{fitness-proportional selection}
\citep{Hof:88}, we adopt it here since the relative performance of a
certain strategy can be also interpreted as its \emph{fitness}.

The dynamics of \eqn{a-dyn} does not take into account that agents can
migrate between the $K$ islands. For simplicity, we assume that each
island $k$ is occupied by the same number of agents $N_{k}=N/K$ and
migration can occur between between any two islands with a constant and
equal migration rate $m_{uv}\equiv m/(K-1)$. We interpret $m \in\{0,1\}$
as the total fraction of the agent population at any given island that
migrates to other islands in a given time interval, namely one
generation. In a first approximation, we assume that migration only
occurs \emph{after} a generation is completed, i.e. at fixed times $G$,
$G+1$.

While the fraction $m$ is fixed (but controlled for, afterwards), the
\emph{composition} of the migrating subpopulation across the existing
strategies $z^{(s)}$ is not. We assume this composition to be proportional to
the fraction $f_{k}^{(s)}$ of these strategies at island $k$ at the time
when generation $G$ is completed and migration occurs, to affect the
dynamics during the next time interval $G+1$. This means migration
changes the fraction $f_{k}^{(s)}(G+1)$ by an additional amount
$\mathcal{F}$ representing the difference between the influx and the
outflux of the agents playing strategy $z^{(s)}$:
\begin{eqnarray}
  \label{eq:migration-only}
  \mathcal{F}_k^{(s)} (G) &=& - m\, f_{k}^{(s)}(G)\; 
  + \; \frac{m}{K-1} \sum_{j=1, j
    \neq k}^K f_{j}^{(s)}(G) 
\end{eqnarray}
which defines the complete dynamics by the following set of iterative
equations:
\begin{equation}
  \label{eq:migration}
  f_k^{(s)} (G+1) = \frac{a_k^{(s)}(G)}{\bar{a}_k(G)} f_k^{(s)}(G) +
  \mathcal{F}_{k}^{(s)}(G)
\end{equation}
Using $\mathcal{F}(0)=0$ and an equal share $f^{s}_{k}(0)$ for each
strategy on each island as initial conditions, the dynamics is completely
determined if we know the respective average payoffs which are derived in
the following section. We note that, in general, the fixpoints of a
difference equation may not characterise its asymptotic behaviour, while
it holds for the given simplified case.

\subsection{Determining the payoffs of repeated interaction}
\label{payoff}

In a deterministic game, we are able to calculate the average payoff
$a_{i}^{(rq)}$ that is received \emph{by agent $i$} playing strategy
$z^{(r)}$, $r \in s$, $n_{g}$ times with an agent $j$ playing strategy
$z^{(q)}$, $q \in s$. For $s=1,2,3,4$, the results are given in the
$4\times 4$ payoff matrix of \eqn{strat-payoff}. Note that the matrix is
not symmetric, since it gives the average payoff of agent $i$.  $r$
refers to the row and $q$ to the column of \eqn{strat-payoff}.
\begin{equation}
  \label{strat-payoff}
  \boldsymbol{a}_{i}= \frac{1}{n_{g}} \left[\begin{array}{cccc}
      n_{g} R & S + (n_{g}-1)P & n_{g} R & a^{(14)} \\
      T + (n_{g}-1) P & n_{g} P & n_{g} T & P+ (n_{g}-1)T \\
      n_{g} R & n_{g} S & n_{g} R & n_{g} S \\
      a^{(41)} & P + (n_{g}-1) S& n_{g}T & a^{(44)}  
    \end{array}\right]
\end{equation}
where
\begin{eqnarray}
  \label{a-qr}
  a^{(14)} &=& (n_{g}\,\mathrm{div}\, 4)\, \mathcal{P} 
  +  \sum_{k=1}^{ (n_{g}\, \mathrm{rem}\, 4)} b_k  \; ; \quad
  a^{(41)} = (n_{g}\,\mathrm{div}\, 4)\, \mathcal{P} 
  +  \sum_{k=1}^{ (n_{g}\, \mathrm{rem}\,4)} c_k 
  \nonumber \\
  a^{(44)} &=& (n_{g}\,\mathrm{div}\, 2)\, (P+R)
  +  \sum_{k=1}^{ (n_{g}\, \mathrm{rem}\,2)} d_k 
\end{eqnarray}
The symbols T, R, P and S refer to the payoff matrix of the 2-person
game, \eqn{2payoff}.  Further, we have used $n_{g}=(n_{g}\,\mathrm{div}\,
y)\,y+(n_{g}\,\mathrm{rem}\, y)$, with $0\leq (n_{g}\,\mathrm{rem}\, y) <
y$.  $(n_{g}\,\mathrm{div}\, y)$ means the \emph{integer} part of the
division $n_{g}/y$, i.e. $(n_{g}\,\mathrm{div}\, 4)=4$ for $n_{g}=18$;
while $(n_{g}\, \mathrm{rem}\,y)$ means the \emph{remainder}, i.e.
$(n_{g}\, \mathrm{rem}\,4)=2$ for $n_{g}=18$. The constants appearing are
defined as follows:
\begin{eqnarray}
  \label{const}
 \mathcal{P}=T+R+P+S \;; \
b_{1}=S;\; b_2=P;\; b_3=T;\; b_4=R \nonumber \\
c_1=T;\; c_2=P;\; c_3=S;\; c_{4}=R \;\ d_{1}=P;\;d_{2}=R
\end{eqnarray}
To verify how the matrix entries of \eqn{strat-payoff} are derived we
have provided a number of illustrative examples shown in \ref{appA}.

With the known average payoff resulting from each possible interaction,
the average payoff per strategy at island $k$ is simply given as:
\begin{equation}
  \label{a-per}
  a_{k}^{(s)} = \sum_{r=1}^{4} a^{(rs)} f_{k}^{(r)} \;; \quad 
  \sum_{r=1}^{4}   f_{k}^{(r)} =1 \quad (s=1,..,4)
\end{equation}
where $f_{k}^{(r)}$ denotes the fraction of agents playing strategy
$z^{(r)}$ at island $k$ and the $a^{(rs)}$ are given in
\eqn{strat-payoff}. Stricly speaking, \eqn{a-per} has to consider the
fact that an agent does not play against itself. This would lead to a
correction term of the order of $a^{(ss)}/N$, which is small and
therefore neglected here.

\subsection{Calculating the fixpoints}
\label{3.1}

With the above specifications we are now able to solve the dynamic
equations with respect to the fraction of agents $f_{k}^{(s)}(t)$ playing
strategy $z^{(s)}$. Here, we are mostly interested in the impact of
migration on the prevailence of certain strategies, therefore we will
discuss first the case \emph{without} migration, \eqn{a-dyn}, which
should be used as a reference for the case \emph{with} migration,
eq. (\ref{eq:migration}). Also, we are mostly interested in the
\emph{stationary} solutions of the respective dynamics, i.e. in the
fixpoints $f_{k}^{(s)}$ reached after a suffiently large number of
generations.  Our focus is then on three issues which crucially determine
the long-term outcome:
\begin{description}
\item (a) the initial frequencies of strategies $f^{(s)}(0)$,
\item (b) the number of repeated interactions, $n_{g}$, as this enters
  the payoff matrix, eq. (\ref{strat-payoff}) and therefore affects the
  average payoff per strategy, eq.
  (\ref{a-per})
\item (c) the impact of the migration rate $m$ on the prevailing
  strategies.
\end{description}
The stationary values of $f_{k}^{(s)}$ are reached if the different
frequencies do not change anymore in the new generation. This leads to
the stationary condition:
\begin{equation}
  \label{stat}
  f^{(s)}_{k}(G+1)=f^{(s)}_{k}(G)=f_{k}^{(s)}
\end{equation}
where $f_{k}^{(s)}$ shall denote the stationary (i.e. asymptotic)
frequency of strategy $z^{(s)}$, in contrast to $f_{k}^{(s)}(0)$ that denotes
the initial frequency.  For the full dynamics which includes migration,
eq. (\ref{eq:migration}), we solve this condition numerically. But for
the case without migration, \eqn{a-dyn}, the stationary condition leads
to $\bar{a}_{k}f_{k}^{(s)}=a_{k}^{(s)}f_{k}^{(s)}$, which reduces to the
quite simple expression $\bar{a}_{k}=a_{k}^{(s)}$ only if
$f_{k}^{(s)}\neq 0$.

To further discuss the case of isolated islands, we drop the index
$k$. For $m=0$ \eqn{stat} with eq. (\ref{a-dyn}) basically yields two
different types of solutions, which could be either stable or unstable:
\begin{itemize}
\item[(i)] the exclusive dominance of any \emph{one} strategy $z^{(q)}$
  $(q \in s)$ while all other strategies $z^{(r)}$, $r\neq q$, have
  disappeared, i.e.
  \begin{equation}
    \label{exclus}
    f^{(q)}=1 ;\; \bar{a}_{k}=a^{(q)}  \quad \mbox{and} \quad 
    f^{(r)}=0 \;\; \mbox{for}\;\; r\neq q 
  \end{equation}
\item[(ii)] the co-existence of some (or all) of the strategies
  $z^{(q)}$, $z^{(r)}$ with the \emph{same average payoff} $\bar{a}$, but
  probably \emph{different frequencies of agents}, $f^{(q)}$, $f^{(r)}$,
  i.e.
  \begin{equation}
    \label{coex}
    \sum_{i=1}^{4} a^{(iq)} f^{(i)} =  \sum_{i=1}^{4} a^{(ir)} f^{(i)}     
  \end{equation}
\end{itemize}
In the deterministic case considered here it solely depends on the
initial conditions $f^{(s)}(G=0)$, which of the possible stationary
solutions will be eventually reached. After we know the \emph{basins
of attraction}, i.e. the range of possible initial conditions that lead to a
given stationary solution and the \emph{separatrix} deviding them, we
could tell from the outset which strategies will be adopted by the
population, and which strategies will disappear.  Therefore, in the next
section we will calculate the basins of attraction for the case of isolated
islands, to compare these later on for the case of migration which, as we
will see, shifts these basins considerably.

\section{No migration: Coexistence and dominance of strategies}
\label{4d}

For isolated islands, condition (\ref{stat}) leads to $s=4$ coupled
equations, which follow from \eqn{a-bar} with \eqn{a-per}. In addition,
we have the boundary condition $\sum_{s} f^{(s)}=1$, which leads to three
independent variables $f^{(1)}$, $f^{(2)}$, $f^{(3)}$. The separatix then
appears as a two-dimensional \emph{plane} in this three-dimensional
space, which will be hard to calculate analytically.  In order to
elucidate this multi-dimensional problem, we discuss in \ref{2d} the case
of only two possible strategies, where the separatrix appears as a
\emph{point} in the one-dimensional space, and in \ref{3d} the case of
three possible strategies, where the separatrix appears as a \emph{line}
in the two-dimensional space. Here, we continue with the full problem of
four possible strategies.

Compared to \ref{3d}, strategy $z^{(4)}=$ A-TFT is added which in the
long run benefits agents playing ALL-D and therefore will decrease the
basins of attraction of the cooperating strategies.  I.e. considering A-TFT as
an additional strategy will create a more difficult environment for the
invasion of cooperation, therefore it will be very interesting to see,
under what conditions cooperation can survive also in the worst case.
The stationary solutions for the frequencies $f^{(s)}$ result from the
following set of coupled equations:
\begin{equation}
  \begin{aligned}
    \label{eq:steady_4}
    \bar{a}=\sum_{s=1}^4 a^{(s)} f^{(s)}=
    \sum_{s=1}^4 \left(\sum_{r=1}^4 a^{(rs)} f^{(r)} \right) f^{(s)} \\
    \bar{a} f^{(i)} - \Big(a^{(i1)}f^{(1)}+a^{(i2)} f^{(2)}
    +a^{(i3)}f^{(3)}+a^{(i4)}f^{(4)}\Big)f^{(i)} &=0 \quad (i=1,2,3,4) \\
    f^{(1)}+f^{(2)}+f^{(3)}+f^{(4)}&=1
  \end{aligned}
\end{equation}
The matrix elements $a^{(rq)}$ can be calculated from
\eqn{strat-payoff}. With $n_{g}=4$ we find the following stationary
solutions:
\begin{equation}
  \label{separat-4}
  \begin{array}{cl}
    (i)\quad & f^{(1)}=1,\; f^{(2)}=0, \; f^{(3)}=0, \; f^{(4)}=0  
    \quad \mbox{(stable)}\\ 
    & f^{(1)}=0,\; f^{(2)}=1,\; f^{(3)}=0, \; f^{(4)}=0 
    \quad \mbox{(stable)}\\
    & f^{(1)}=0,\; f^{(2)}=0,\; f^{(3)}=1, \; f^{(4)}=0 
    \quad \mbox{(unstable)}\\
    & f^{(1)}=0,\; f^{(2)}=0,\; f^{(3)}=0, \; f^{(4)}=1 
    \quad \mbox{(unstable)}\\
    (ii)\quad & f^{(1)}_{\mathrm{thr}}=0.2,\; f^{(2)}_{\mathrm{thr}}=0.8,\;
    f^{(3)}=0 , \; f^{(4)}=0 
    \quad (\mbox{for} \;\;n_{g}=4)\quad \mbox{(unstable)}\\
    & f^{(1)}=1.0-f^{(3)}\; (0\leq
    f^{(3)}\leq 1),\; f^{(2)}=0 , \; f^{(4)}=0 \quad \mbox{(stable)}
  \end{array}
\end{equation}
We note that the stationary solutions also cover the results of the
two-strategy case, \ref{2d} and the three-strategy case, \ref{3d}. We
find solutions of type (i) where one strategy is exclusively dominating
the whole population. However, the solution of all agents playing
$z^{(4)}=$ A-TFT is also an unstable attractor, because a small
perturbation by agents playing either strategy ALL-D or TFT will lead to
the invasion of that respective strategy into the whole A-TFT population.
To see this, we also refer to the payoffs received in the respective
interactions, \eqs{4-2}{1-4}.  From the solutions (ii) that describe the
coexistence of strategies we further see that the addition of strategy
$z^{(4)}=$ A-TFT does not lead to new coexisting states compared to the
case of two strategies, \ref{2d} and three strategies, \ref{3d}. There is
only a coexistence between strategies $z^{(1)}=$ TFT and $z^{(3)}=$
ALL-C, which is stable as long as no spontaneous mutations towards ALL-D
or A-TFT occur.

The separatrix that distinguishes between the different attractors is in
the given case a two-dimensional plane in the three-dimensional space of
independent frequencies with $\sum_{s}f^{(s)}=1$. With respect to the
cooperative strategy TFT it defines the \emph{threshold frequency}
$f_{\rm thr}^{(1)}=f^{(1)}(G=0)$ that has to be reached initially, in
order to allow the system to converge to a fully cooperative state,
$f^{(1)}(G\to \infty)=1$.  Because of the $n_{g}$-dependence of the
payoff matrix, \eqs{strat-payoff}{matrix-2} $f^{(1)}_{\mathrm{thr}}$
strongly decreases with the number of encounters $n_{g}$, as shown in
\pic{fc-3}. In agreement with known results from evolutionary game theory
\citep{Fogel:95, schweitzer02:_evolut}, for a sufficiently large number
of interactions the whole population will adopt the cooperative strategy,
this way increasing the average payoff.  The important point to notice
here is the quantitative analysis which allows us to determine, for a
given \emph{finite} number of $n_{g}$, the threshold initial frequency
$f^{(1)}_{\mathrm{thr}}$ of agents playing TFT, in order let cooperation
invade the whole population. Or else, for a given initial frequency
$f^{(1)}(0)$, we are able to determine the \emph{minimum number of
  interactions}, $n_{g}^{\mathrm{min}}$, to let the whole population
adopt the TFT strategy.

\begin{figure}[htbp]
  \centerline{\includegraphics[width=6.0cm]{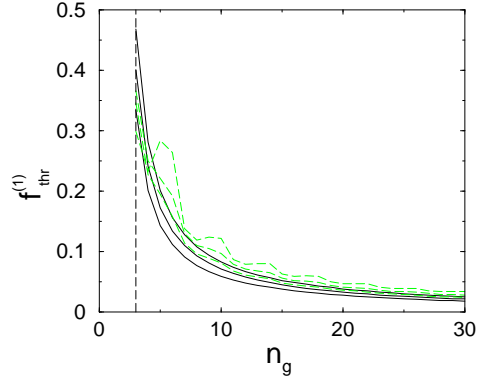}}
  \caption{ Threshold frequency $f^{(1)}_{\mathrm{thr}}=f^{(1)}(0)$ of
    agents playing strategy $z^{(1)}=$ TFT vs. number of interactions
    $n_{g}$ between each two agents, in the case of four strategies.  For
    $f^{(1)}(0)\geq f^{(1)}_{\mathrm{thr}}$, the TFT strategy will be
    adopted by the whole population, while for $f^{(1)}(0)\leq
    f^{(1)}_{\mathrm{thr}}$ the $z^{(2)}=$ ALL-D strategy will be adopted
    by the whole population.
      \label{fc-3} }
\end{figure}

To get some further insights, we find it more convenient to present
two-dimensional graphical projections of this three-dimensional space,
where the separatrix naturally appears as a line. The most interesting
projection is the $f^{(1)}-f^{(2)}$ projection, since it allows us to
compare the results with the previous cases.  The results are shown
\pic{fig:basin_4}. They have been obtained by numerically solving
\eqn{a-dyn} for the full range of initial frequencies, $0\leq
f^{(s)}(0)\leq 1$. In the stationary limit, the average payoff was then
evaluated to find out to which basin of attraction the solution belongs.
\begin{figure}[htbp]
  \centerline{\includegraphics[width=5.0cm]{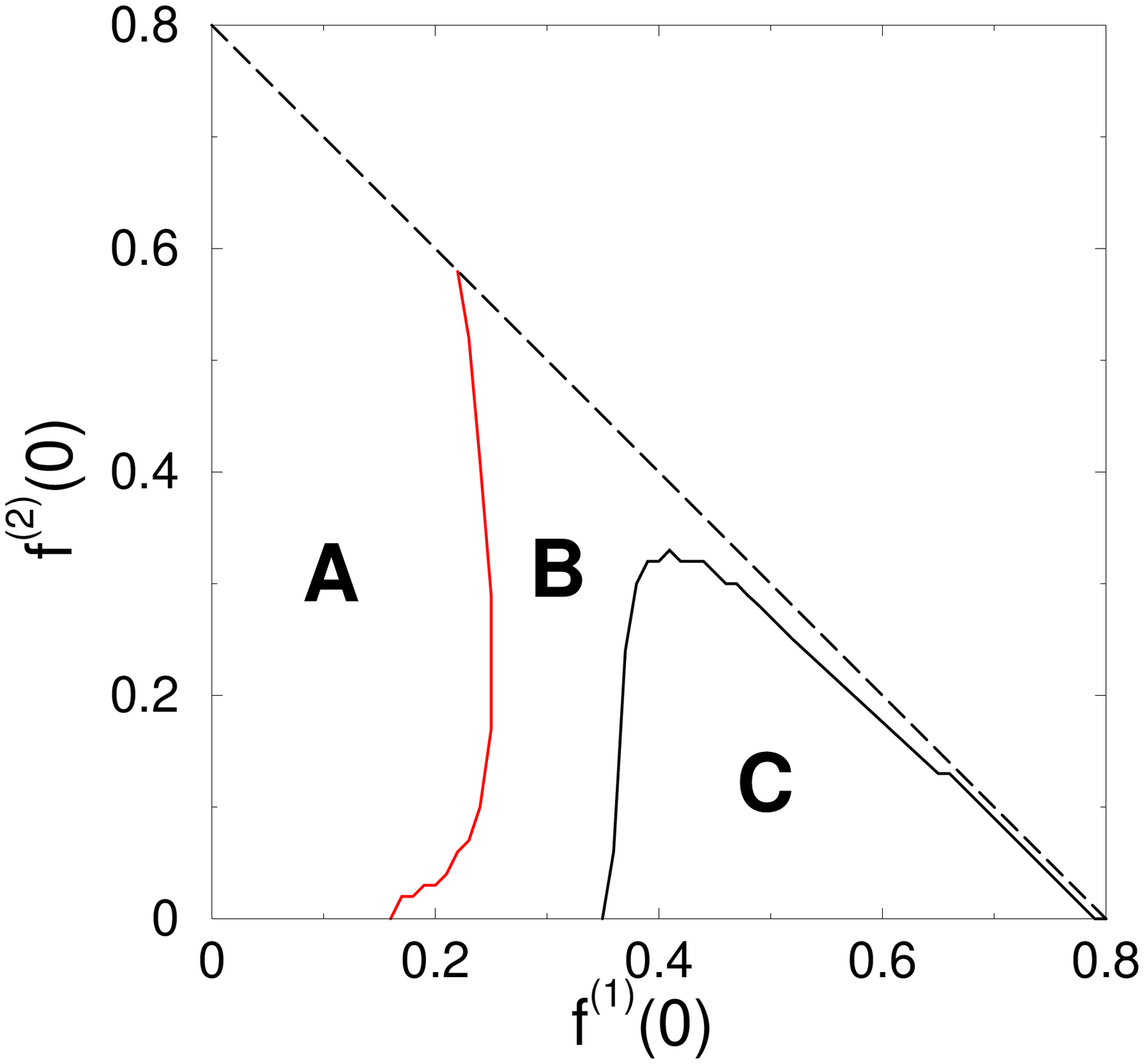}\hfill
    \includegraphics[width=5.0cm]{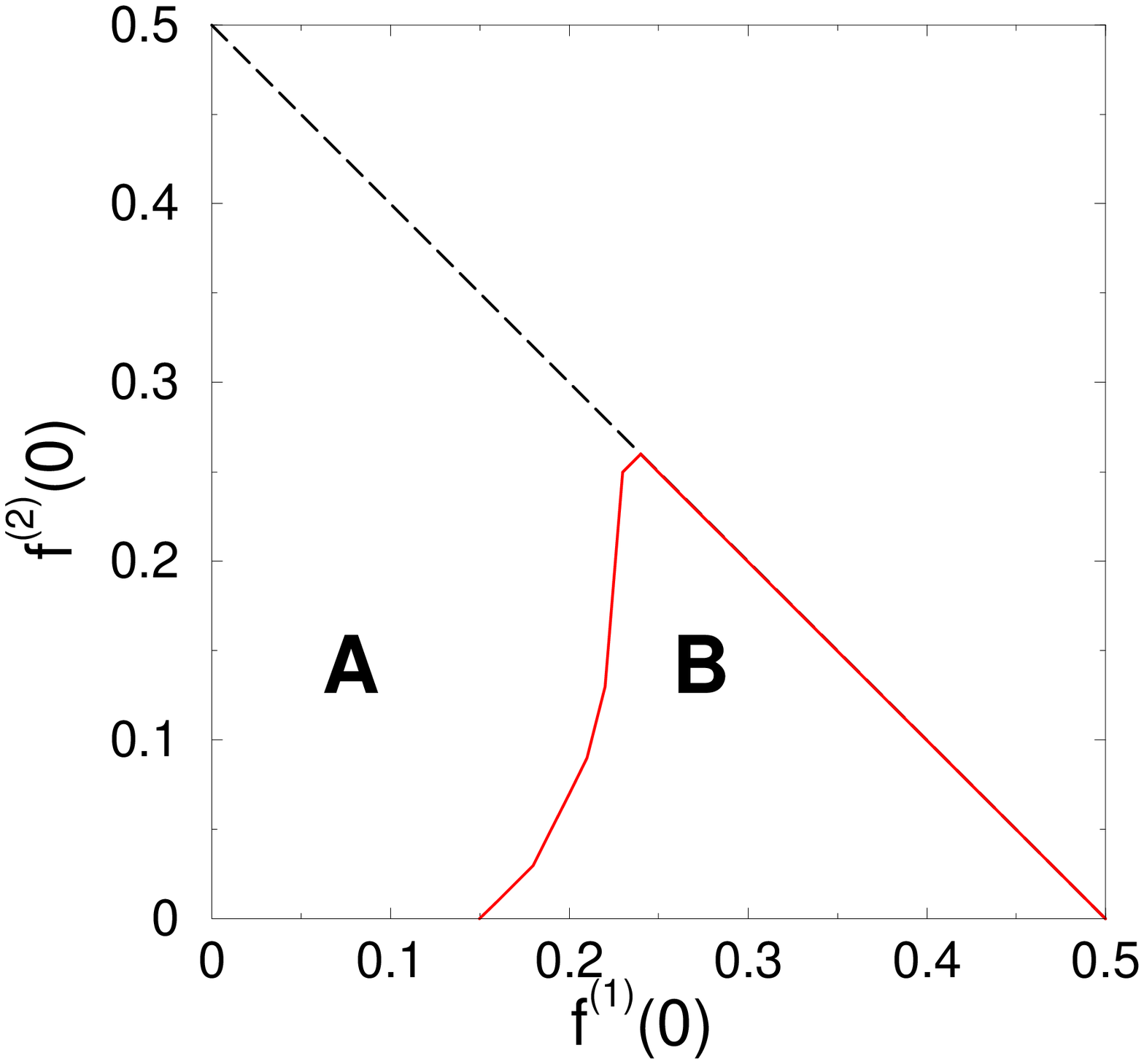}\hfill
    \includegraphics[width=5.0cm]{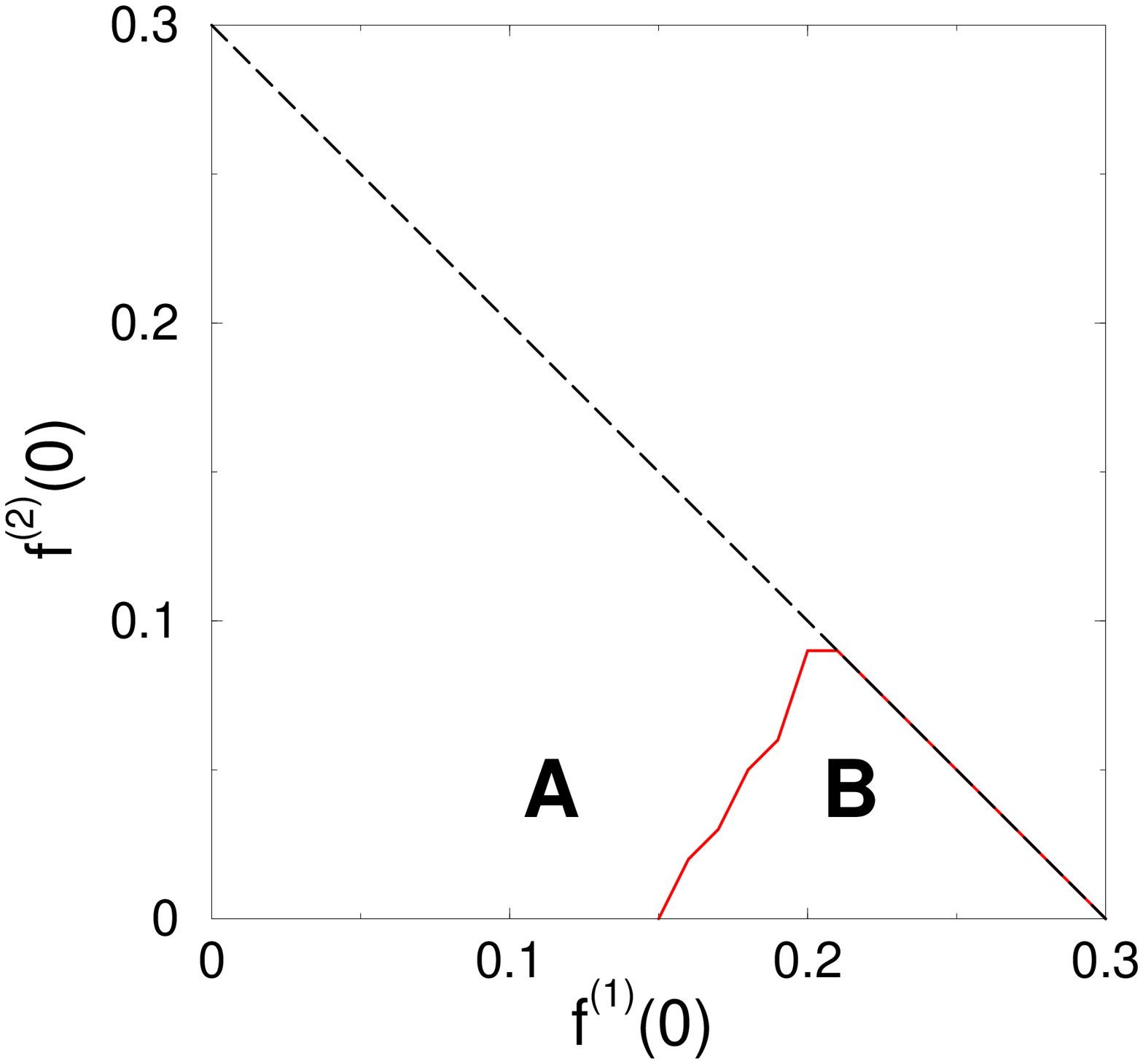}}
\caption{Basins of attraction, i.e. range of initial frequencies
  $f^{(s)}(0)$ that lead to a particluar stable solution,
  \eqn{separat-4}, in the case of four strategies.  $A$: adoption of
  $z^{(2)}=$ ALL-D strategy in the whole population, $B$: adoption of
  $z^{(1)}=$ TFT strategy in the whole population, $C$: coexistence of
  both $z^{(1)}=$ TFT and $z^{(3)}=$ ALL-C strategies.
  $f^{(3)}(0)=1-f^{(1)}(0)-f^{(2)}(0)$, $n_g=4$.  (left)
  $f^{(4)}(0)=0.2$, (middle) $f^{(4)}(0)=0.5$, (right) $f^{(4)}(0)=0.7$.
  \label{fig:basin_4}}
\end{figure}

The calculation of the basins of attraction in \pic{fig:basin_4} are done
for a fixed number of interactions, $n_{g}=4$, but for a varying initial
frequency of agents playing strategy $z^{(4)}=$ A-TFT. Of course, this
sets limits to the \emph{maximum} initial frequencies
$f^{(1)}(0)=1-f^{(3)}(0)$ and $f^{(2)}(0)=1-f^{(3)}(0)$, therefore in
\pic{fig:basin_4} we have scaled the $f^{(s)}(0)$-axes by the maximum
possible value, which should be noticed when comparing the figures.  For
a small initial frequency of A-TFT (\pic{fig:basin_4} left) we find the
picture rather similar to the case of three strategies shown in \ref{3d},
however the separation line between the basins of attraction $A$
(defection) and $B+C$ (cooperation) is now a non-linear function,
different from \eqn{line}.

With an increasing initial frequency of A-TFT, we find that the basin of
attraction of defection is naturally decreasing in absolute size, but it
is increasing \emph{relative} to the basin of attraction of
cooperation. That means that an increase in the fraction of agents
initially playing A-TFT always benefits the adoption of the ALL-D
strategy, at the end. A similar conclusion can be drawn also for the
number of agents initially playing ALL-C.  We notice that with an
increasing initial frequency of A-TFT region $C$, that describes the
coexistence of both strategies TFT and ALL-C, ceases to exist,
i.e. strategy ALL-C has vanished for the benefit of strategy ALL-D.  This
can also be understood by looking at the average payoff matrix,
\eqn{strat-payoff}. If an agent playing ALL-D meets with an agent playing
ALL-C, it gets the maximum payoff, $T$, while it gets almost the maximum
payoff $(P+(n_{g}-1)T)/n_{g}$ when it meets an agent playing A-TFT.
Thus, strategy ALL-D will eventually invade the population of agents
playing either ALL-C or A-TFT, or, in other words, the presence of
strategies like ALL-C and A-TFT in the population helps strategy ALL-D to
invade.

Eventually, we have also calculated numerically the relative size of each
basin of attraction as shown in \pic{fig:bar_4} for two different values
of $n_{g}=4$ and $n_{g}\to \infty$. Here the relative sizes $a$, $b$, $c$
and $d$ refer to the three-dimensional space of all possible initial
conditions (which is different from \pic{fig:basin_3} where they refer to
the areas $A$, $B$ and $C$ shown, thus \eqn{area} does not apply here).
When $n_g=4$, the ALL-D basin 'a' is slightly bigger than the cooperative
basin 'd' as shown in contrast to figure \ref{fig:bar_3}. With $n_{g}\to
\infty$, the size of the ALL-D basin 'a' becomes pretty small as compared
to cooperative basin 'd'.

\begin{figure}[htbp]
  \centerline{\includegraphics[width=6.0cm]{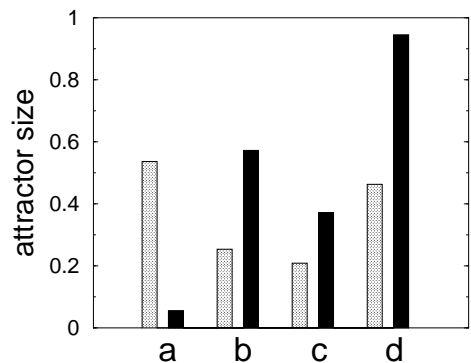}}
\caption{Relative size of the basins of attraction. 'a' represents the size of the
  ALL-D basin, 'b' represents the size of the TFT basin, 'c' represents
  the size of TFT+ALL-C basin and 'd' represents the size of cooperative
  basin (i.e. b+c). The left bars (shaded area) refer to
  $n_{g}=4$, while the right bars (black area)
  refer to $n_{g}\to \infty$. Thus, the change
  indicates the influence of $n_{g}$ on the size of the basins of attraction.
  \label{fig:bar_4}
}
\end{figure}

To conclude this analysis, we have shown for the case of isolated islands
(no migration) what are the critical initial conditions for cooperative
strategies, such as ALL-C, TFT to be adopted by the whole agent
population. Dependent on the number of interactions $n_{g}$, we can tell
the minimum initial frequency of TFT, $f^{(1)}(0)$ that has to be reached
to ensure an cooperative outcome dependent on the initial mix of
strategies (details of the two and three strategy case are provided in
\ref{2d}, \ref{3d}). In the limit of small $n_{g}$, while one may naively
assume that an initial increase of the ALL-C strategy benefits
cooperation, at the end, this analysis shows that with the involvement of
A-TFT it eventually benefits defection in the system, which is an
interesting finding by itself. In the following, we will investigate how
this picture changes if we add the possibility of migration.

\section{Migration: The rise and fall of cooperation}
\label{sec:island}

\subsection{The role of migration}
\label{sec:mig2}

The previous section has shown that an increase of the number of
interaction $n_{g}$ always supports the prevailence of cooperating
strategies, such as ALL-C and TFT. Therefore, in the following we will
concentrate the analysis on the critical conditions which occur for small
$n_{g}$. We fix $n_{g}=4$ (note that $n_{g}=3$ is the threshold number of
interaction to allow the replication of cooperating strategies at all,
see \ref{2d} and Figure \ref{fc-3}.
\begin{figure}[htbp]
  \begin{center}
    \includegraphics[width=0.8\textwidth]{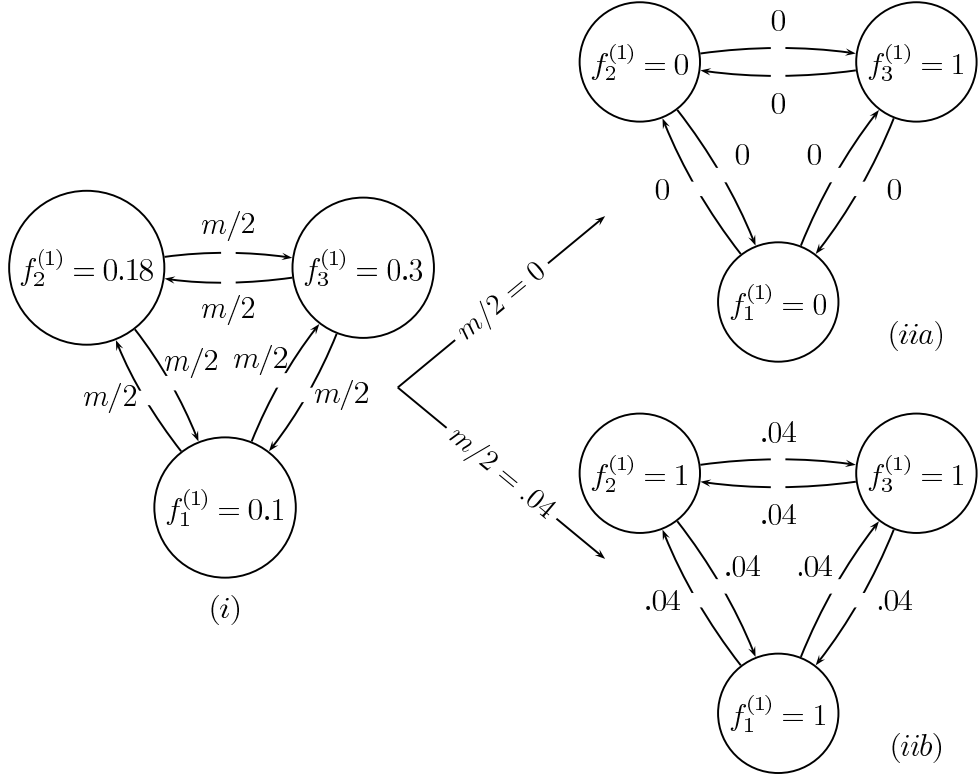}
  \end{center}
  \caption{Migration between the three islands enhances the outbreak of
    cooperation. (i) Initial state ($G=0$) using two strategies. The
    defective strategy ALL-D has the majority on all islands. 
    (ii) Final state ($G\to \infty$) (a) without and (b) with a small
    migration rate. In (a) only one island has adopted cooperation, in
    (b) all three islands have adopted cooperation
  }
  \label{fig:island}
\end{figure}

To understand how migration affects the existing equilibrium states, let
us start with the simple example shown in Fig. \ref{fig:island}. We take
the setup of Fig. \ref{fig:lattice} and consider only two strategies,
$z^{(1)}=$TFT and $z^{(2)}=$ALL-D, but different initial conditions on
the 3 islands. As the detailed analysis of \ref{2d} has shown, below an
initial frequency $f^{(1)}(0)=0.2$ the only stable stationary state is
$f^{(1)}=0$, $f^{(2)}=1$. Hence, if no migration is possible, with the
given initial conditions shown in Fig. \ref{fig:island}(i) the three
islands converge into a final state shown in Fig. \ref{fig:island}(iia)
where ALL-D prevails in two islands, while in one TFT dominates because
that island started from an initial condition above the threshold. If we
however introduce a rather small migration rate of $m/2=0.04$ all three
islands are eventually dominated by the cooperating strategy, TFT, as
shown in Fig. \ref{fig:island}(iib). Hence, even with a rather small
advantage on \emph{one} island ($f^{(1)}(0)=0.3$ is just above the
threshold of 0.2), a small migration was able to induce a transition
toward TFT instead of a relaxation into ALL-D on \emph{all} islands.

It is precisely this kind of phenomenon that we would like to understand
better regarding the critical conditions involved. In particular, we
concentrate on 
the impact of the migration rate on the threshold value of TFT, to still
observe the outbreak of cooperation.  The results are found numerically,
by iterating the set of eqs. (\ref{eq:migration}), for $n_{g}=4$ and $K$
islands that have the same total number of agents, but different initial
strategy distributions.

\begin{figure}[htbp]
\centerline{
    \includegraphics[width=5.3cm]{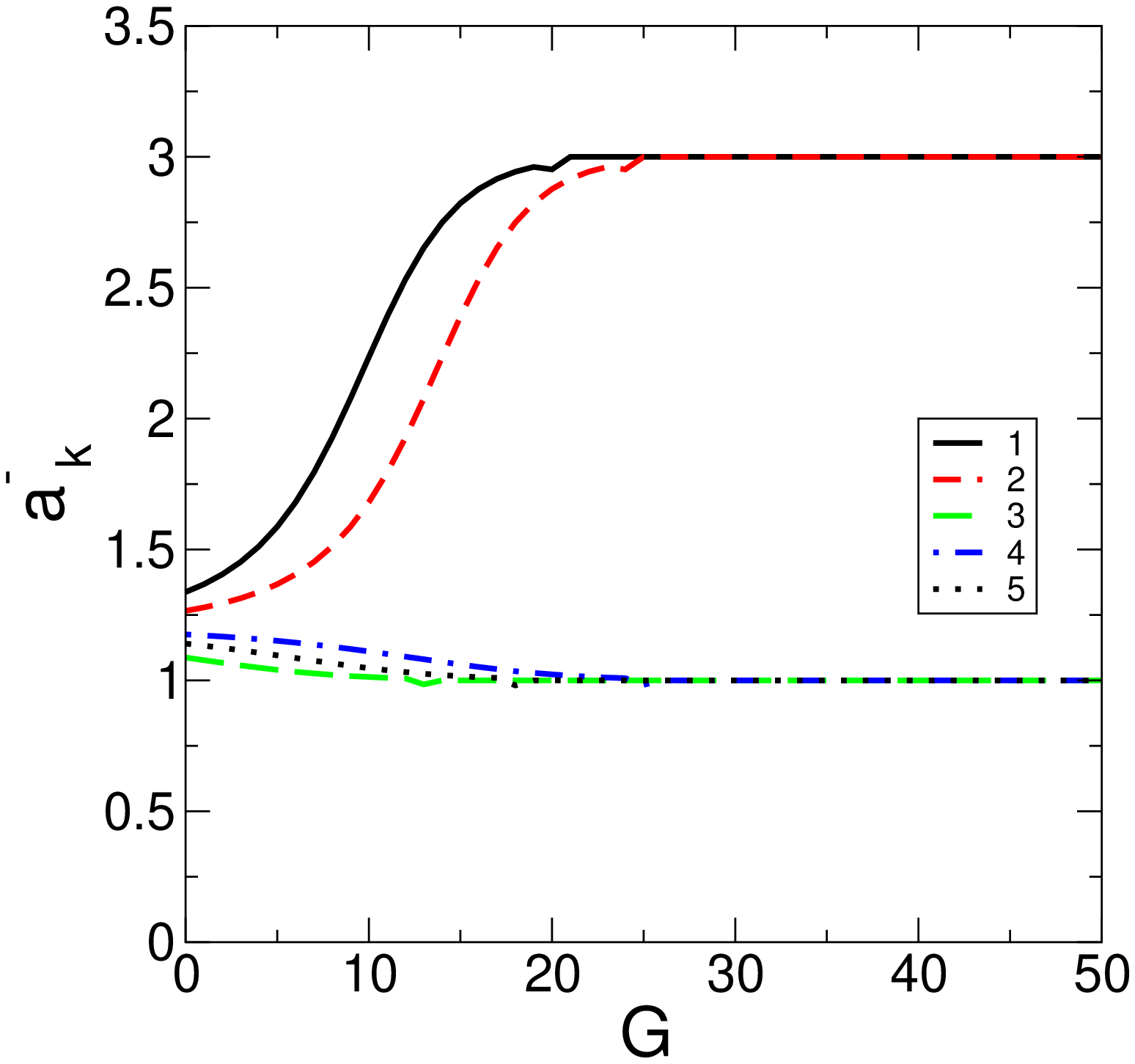}\hfill
    \includegraphics[width=5.3cm]{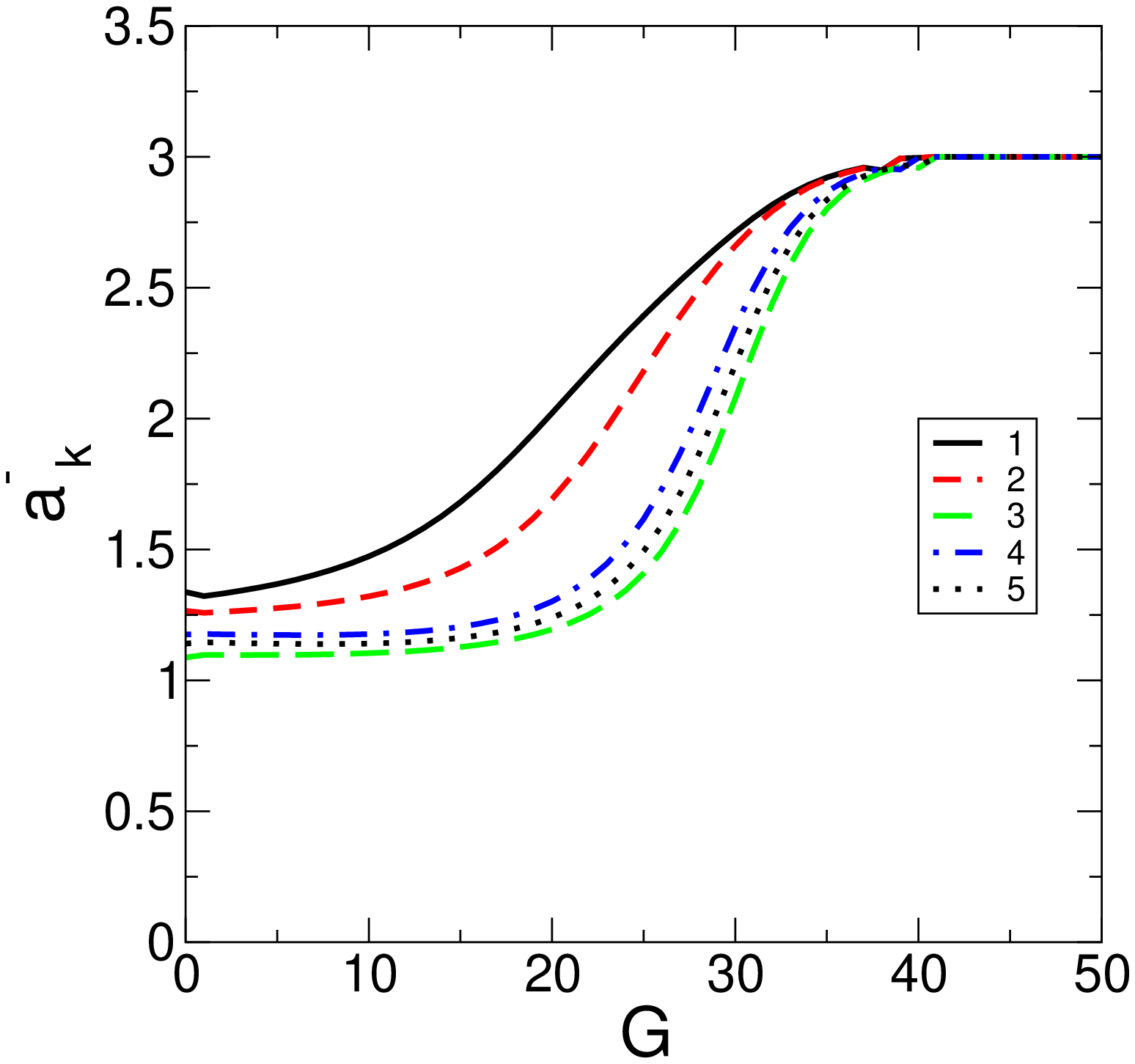}\hfill
    \includegraphics[width=5.3cm]{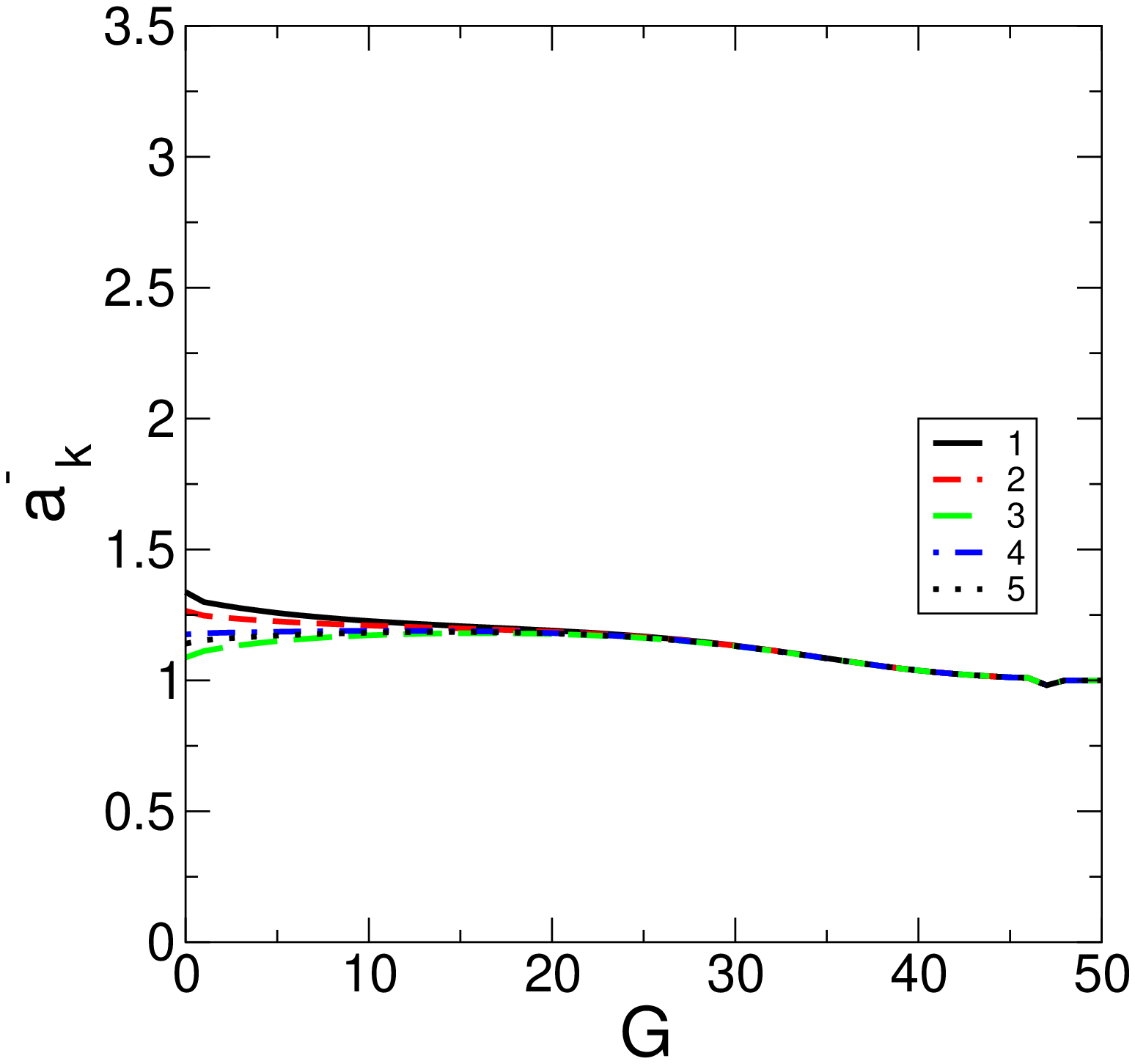}} 
  \caption{Evolution of the average payoff in the $K=5$ island model for
    different migration rates $m$: (left) $m=0.0$, (middle) $m=0.08$,
    (right) $m=0.2$. Parameters: Panmictic population in $K=5$ island
    model consisting of TFT and ALL-D. Initial frequencies of TFT:
    $f_{1}^{(1)}=0.3 , f_{2}^{(1)}=0.25 , f_{3}^{(1)}=0.1 ,
    f_{4}^{(1)}=0.18 , f_{5}^{(1)}=0.15$ and $f_k^{(2)}=1-f_{k}^{(1)}$.}
  \label{fig:evol_2d}
\end{figure}
\pic{fig:evol_2d} shows results of the two strategies case, TFT and
ALL-D, for $K=5$ and three different migration rates, in terms of the
average payoff $\bar{a}_k$. As explained in \ref{2d} $\bar{a}_k=1$ means
that ALL-D invades the entire population, whereas $\bar{a}_k=3$ is
achieved when TFT takes over.  \pic{fig:evol_2d}(left) with $m=0$ is used
for comparison only, as it shows the result predicted in \ref{2d}.  I.e.,
three islands which have a frequency of TFT below the threshold value of
$0.2$ are taken over by agents playing ALL-D. The remaining two islands
are invaded by TFT. This result changes in \pic{fig:evol_2d}(middle),
where we allow migration of $m=0.08$ between islands. Then all islands
become populated solely by TFT which is indicated by $\bar{a}_k=3$.
However, if we further increase the migration rate to $m=0.2$, the
\emph{opposite} result occurs. In this case ALL-D invades all islands as
shown in \pic{fig:evol_2d}(right) where $\bar{a}_k=1$. These results
suggest that there exists \emph{an optimal range for $m$} that ensures
the outbreak of cooperation, which will be further investigated in the
following sections.

\subsection{How migration influences cooperation}
\label{threshold}

\begin{figure}[htbp]
  \begin{center}
    \includegraphics[width=5.1cm]{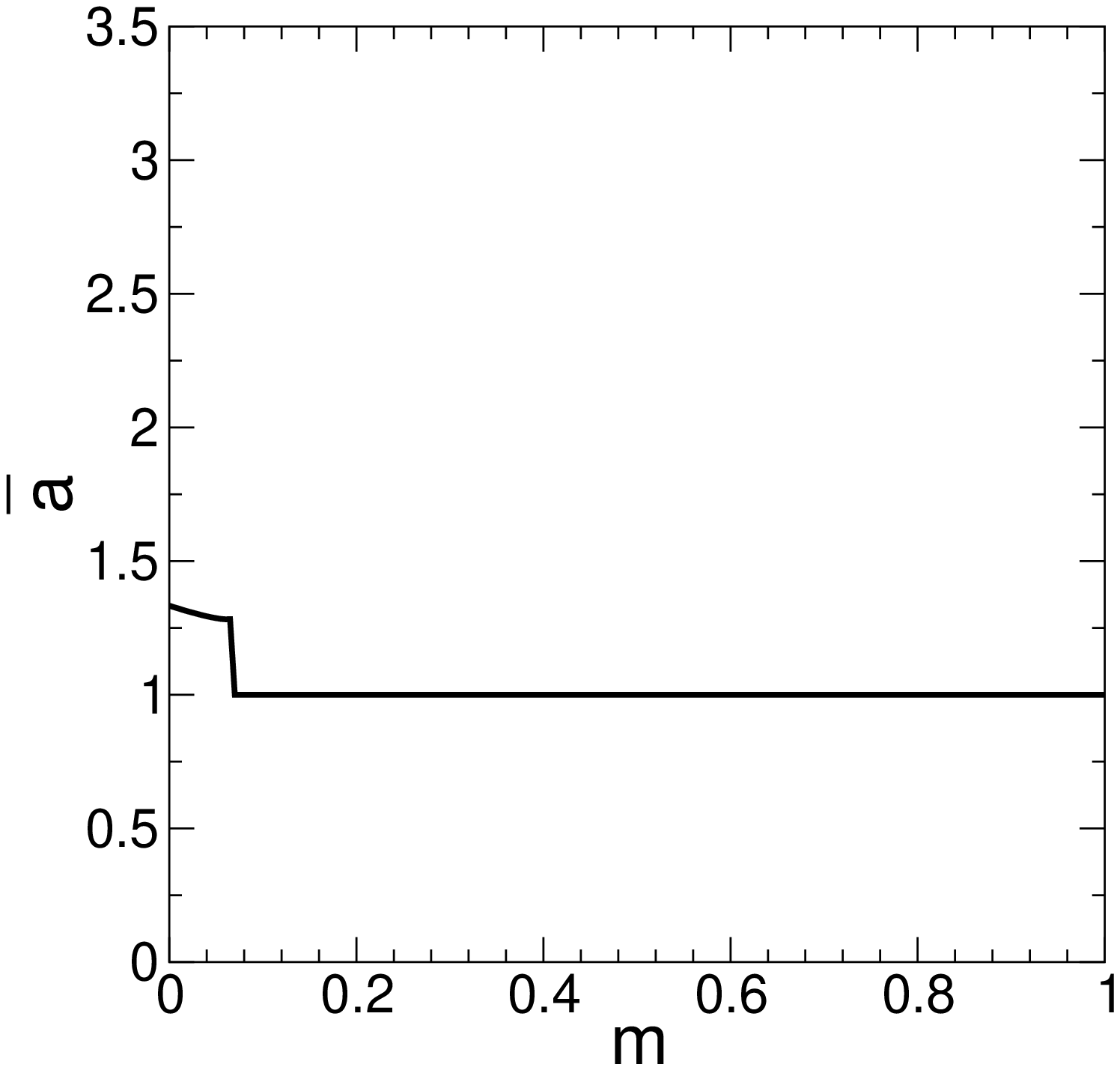} 
\hfill
    \includegraphics[width=5.1cm]{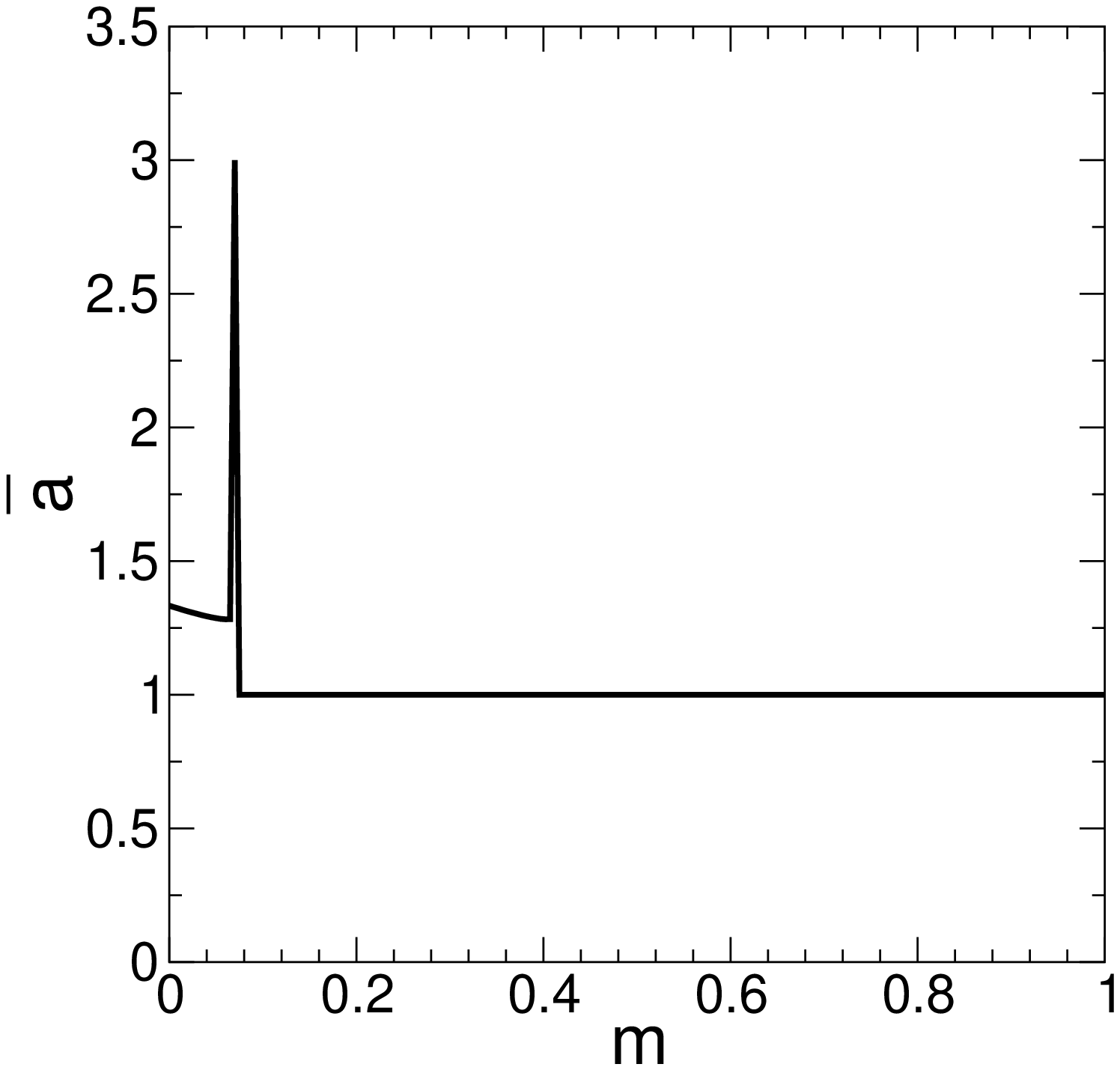} 
\hfill
    \includegraphics[width=5.1cm]{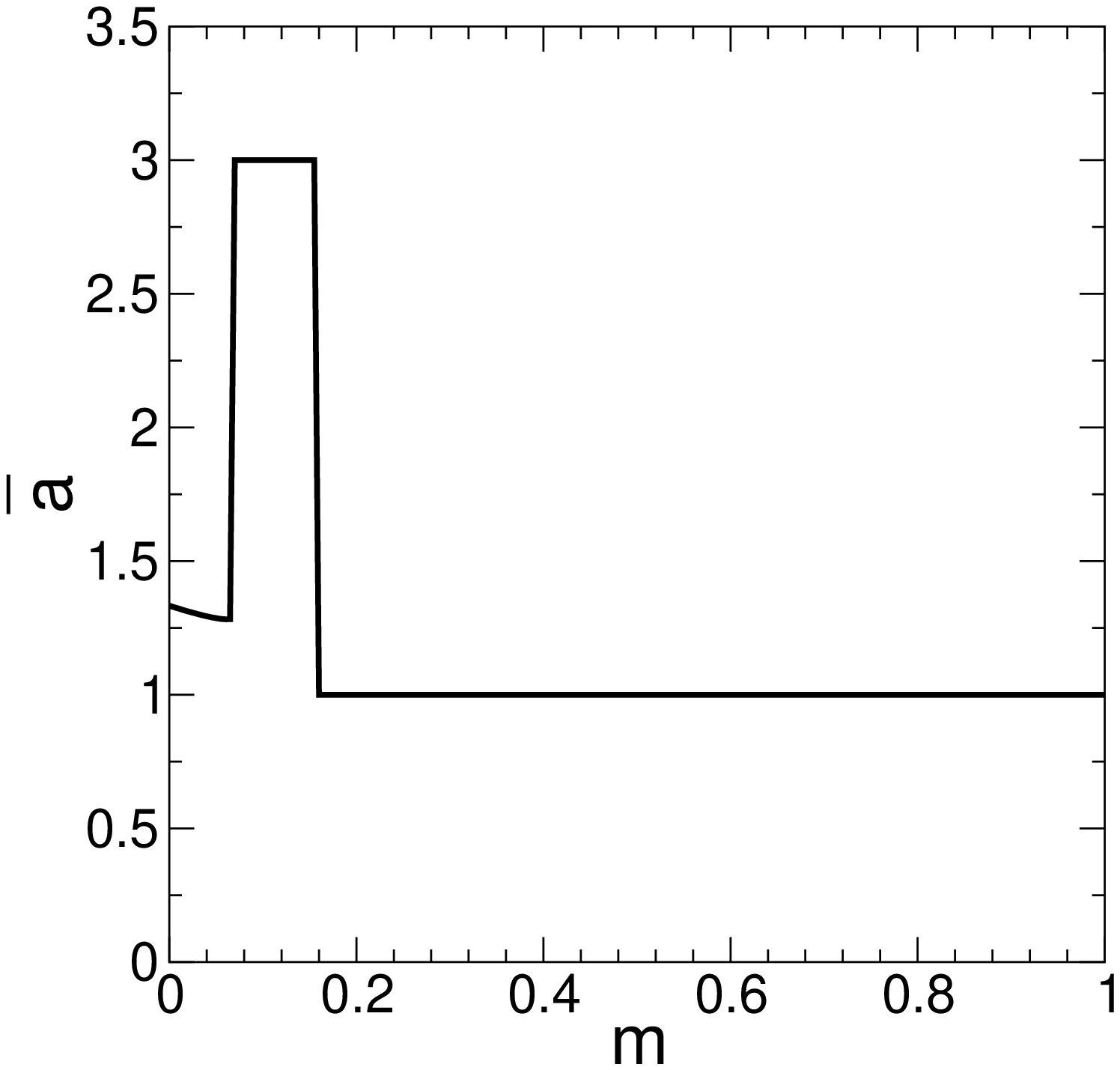}
  \end{center}
  \caption{Increasing $f_1^{(1)}(G=0)$ leads to increasing bandwidth of
    $m$ in which cooperation will emerge in the system. From left to
    right: (1) $f_1^{(1)}(G=0)=0.31$ (2) $f_1^{(1)}(G=0)=0.32$ (3)
    $f_1^{(1)}(G=0)=0.40$ (4) $f_1^{(1)}(G=0)=1.00$. Parameters: $n_g=4$,
    $K=6$}
  \label{fig:howto_find_threshold}
\end{figure}

We continue our discussion for the case of two strategies, TFT and ALL-D,
and now assume that for $K=6$ $(K-1)$ islands are initially dominated by
stategy ALL-D, i.e. $f_k^{(1)}(0)=0$ for $k\in 2,...,6$, while the
\emph{first} island $k=1$ is initially populated by both TFT and ALL-D,
i.e. $0 <f_{1}^{(1)}(0)<1$. We then vary the migration rate $m$ for
different initial frequencies $f_{1}^{(1)}(0)$ to study numerically under
which condition the outbreak of cooperation on all islands can be
obtained.  The results are shown in Fig.  \ref{fig:howto_find_threshold}.
For very small $m$, the average payoff $\bar{a}$ just reflects the mix of
the initial strategies. But, as the middle part of
Fig. \ref{fig:howto_find_threshold} shows, for a certain value of the
initial frequency $f_{1}^{(1)}(0)$, we find a spike $\bar{a}=3$ at a
particular value of $m$ that indicates the \emph{outbreak of
  cooperation}, thanks to an \emph{optimal migration rate}. One would
assume that increasing the initial fraction of TFT would further boost
cooperation but, as Fig. \ref{fig:howto_find_threshold} (right) clearly
shows, the ``bandwidth'' of optimal migration rates $m$ is rather
limited. Precisely, if $f_1^{(1)}(0)=0.40$ the bandwidth of optimal
migration rates is $0.07\leq m \leq 0.095$ which is to be compared to the
maximum bandwidth $0.07\leq m \leq 0.155$ for $f_1^{(1)}(0)=1$.  This
makes sense because a large migration rate may have a contrary effect:
While cooperative agents are ``exported'' to other islands, the same
number of defective agents are ``imported'', which may reduce the
fraction of TFT below the critical threshold. Hence, there is a
nonmonotonous relation between $m$ and the outbreak of cooperation.

\subsection{Calculating the threshold frequency}
\label{sec:threshold}

We are now interested in finding the \emph{minimum inital frequency}
$f_1^{(1)}(0)$ that will lead to invasion of cooperation on all $K-1$
islands, when varying the migration rate $m$.  We define this as the
\emph{threshold frequency}, $f_{\rm thr}^{K}$:
\begin{eqnarray}
  \label{eq:threshold}
  f_{\rm thr}^K(m^{K}) &=&\min_{f_1^{(1)}} 
f_{1}^{(1)}(G=0) 
\\
\mathrm{with} && f_k^{(1)}(G=0)=0 \ \forall \ k \in 2,...,K\;; \quad  
  f_k^{(1)}(G=\infty)= 1 \  \forall \ k \in K   \nonumber 
\end{eqnarray}
For each $f_{\rm thr}^K$, there exists a specific value of the migration
rate $m^{K}\in [0,1]$ such that the outbreak of cooperation happens on
all islands.  $m^{K}$ denotes the the smallest value of the bandwidth of
optimal migration rates shown in Fig. \ref{fig:howto_find_threshold}.
I.e., for fixed $n_{g}$ and a given number of islands $K$ with the
initial conditions specified in eq. (\ref{eq:threshold}), there exist a
tuple of critical values $(f_{\rm thr}^{K},m^{K})$ which determine the
\emph{outbreak of cooperation}. The outbreak shown in
Fig. \ref{fig:howto_find_threshold}(middle) for $K=6$ is observed for $(f_{\rm
  thr}^{K},m^{K})=(0.32, 0.07)$.  This leads us to the question, how the
threshold value and the critical migration rate change if we change the
number of islands.

\begin{figure}[htbp]
  \centerline{
    \includegraphics[width=6.0cm]{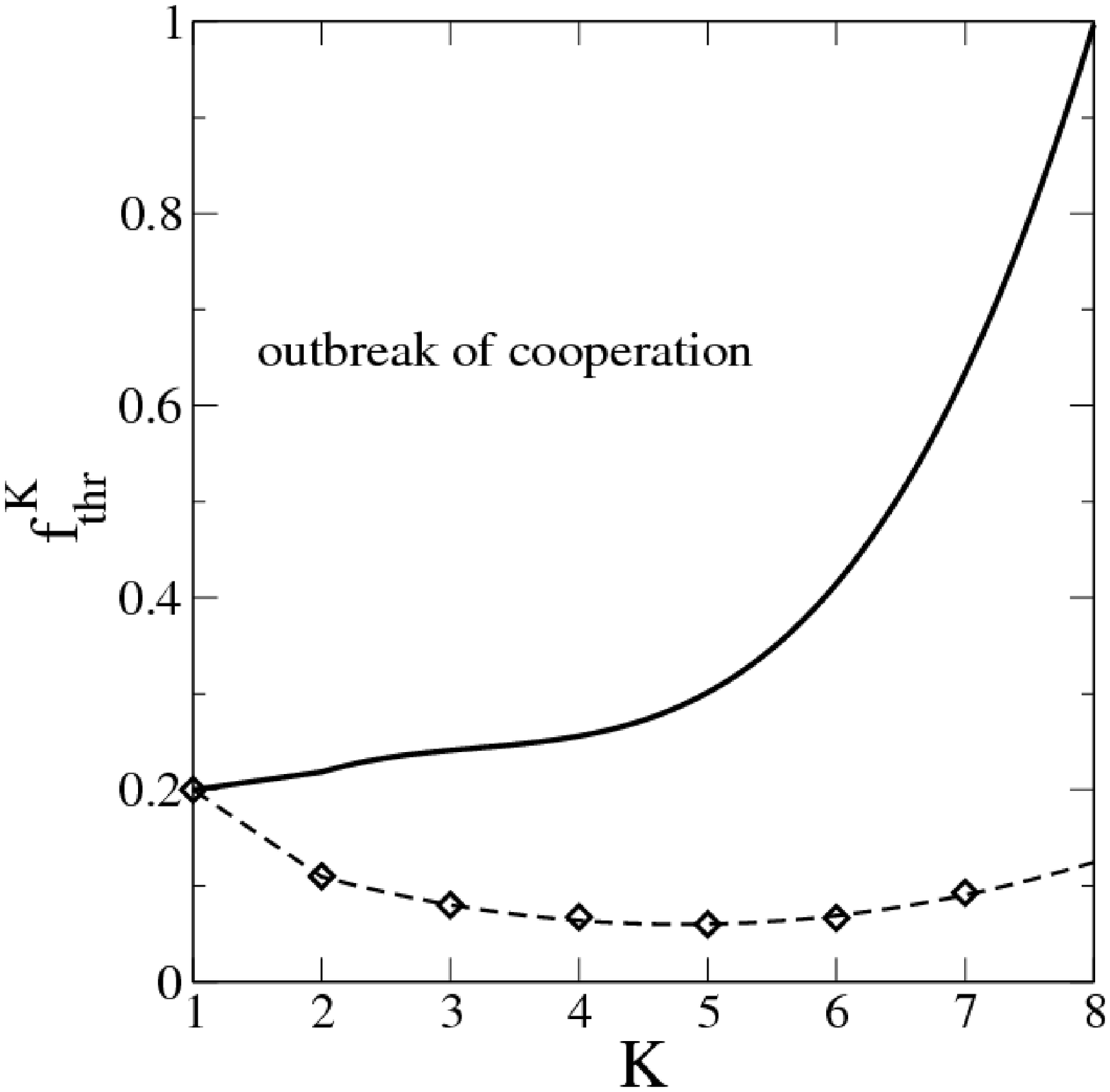} 
\hspace{1cm}
    \includegraphics[width=6.0cm]{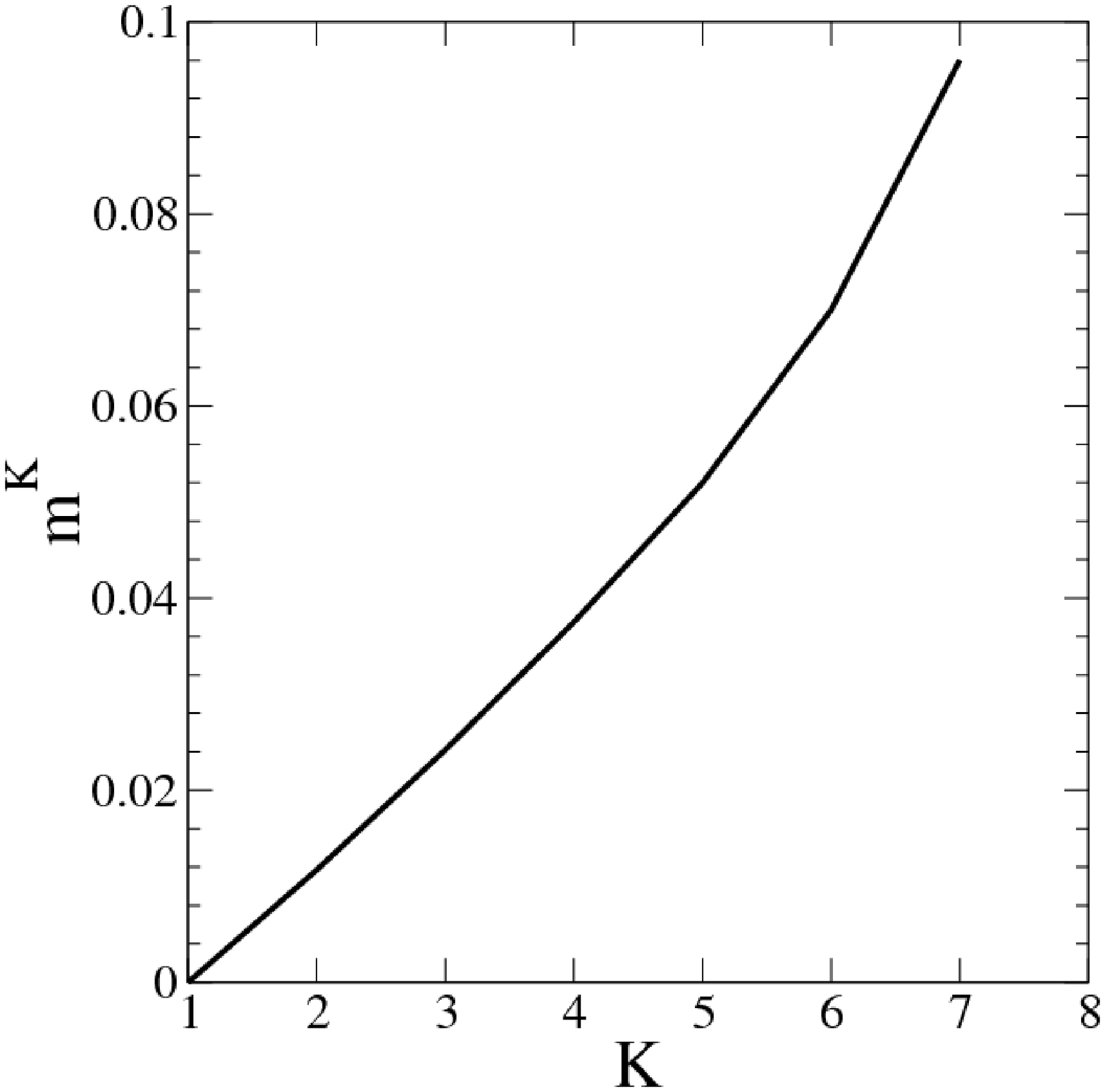}
  }
  \caption{(left) Threshold frequency $f_{\rm thr}^K$ (black line) and
    relative effort $f_{\rm thr}^K/K$ (dashed line) dependent on the
    number of islands $K$. Symbols represent results of numerical
    calculations. Fitted values of eq. (\ref{eq:nonlin-k})
$A_{0} = 0.006$, $A_{1} = 4.798$, $A_{2} = 0.060$. (right) Optimal
migration rate $m^K$ dependent on the  number of islands $K$.
\label{fig:migra_vs_k_and_thres_vs_k}}
\end{figure}
The results of exhaustive numerical calculations which had to search for
the outbreak of cooperation on the whole $(K,f_{1}^{(1)}(0),m)$ parameter
space are shown in \pic{fig:migra_vs_k_and_thres_vs_k}(left). Obviously, for
$K=1$ the threshold frequency results from the analysis given in
\ref{2d}. If the number of islands increases, $f_{1}^{(1)}(0)$ has to
increase as well, in order to be able to ``export'' cooperating agents
while still staying above the theshold. This increase, however, is
nonlinear in $K$. Precisely, as can be verified in
\pic{fig:migra_vs_k_and_thres_vs_k},
\begin{eqnarray}
  \label{eq:nonlin-k}
  f_{\rm thr}^{K}= K^{3} \, A_{0}- K^{2}\, (2 A_{0}A_{1}) + K\, (A_{0}A_{1}^{2}+A_{2})
\end{eqnarray}
The values of the constants $A_{n}$ depend implicitely also on the payoff
matrix, eq. (\ref{2payoff}), which defines the comparative advantage of
each strategy for reproduction and invasion, subsequently. Plotting
$f_{\rm thr}^{K}/K$ we see that for $2\leq K \leq 6$, with every new
island added the \emph{relative effort} to invade the new one with
cooperators \emph{decreases}, while for $K\geq 6$ this effort
\emph{increases} quadratically with $K$. However, the effort of invading
other islands with cooperators always stays below the threshold of an
isolated island, which is 0.2. This is a direct consequence of the
combined processes of migration and reproduction. Newly arriving
cooperative agents will reproduce on the other islands, this way reaching
the threshold value the before the next generation is ready for
migration. Hence, \emph{diversification of the reproduction sites} for
cooperating agents lower the relative effort for the outbreak of
cooperation and makes it more favorable.

As a second consequence of the nonlinear dependency, the
export of cooperation from \emph{one} island to $K-1$ other islands is
only possible for a limited number of islands, because of the two effects
already mentioned above. I.e. the ``export'' of cooperating agents and
the ``import'' of defecting agents both decrease the fraction of
cooperating agents below the threshold frequency. Hence, for the two
strategy case discussed here, $K=8$ already leads to the collapse of
exporting cooperation.
 
The critical migration rate $m^{K}$ dependent on the number of islands is
shown in \pic{fig:migra_vs_k_and_thres_vs_k}(right). We recall that this
gives the minimal fraction of the population that has to migrate in order
to allow the outbreak of cooperation on any island. If
$f_{1}^{(1)}(0)=f_{\rm thr}^{K}$ it is at the same the maximum fraction
to not let cooperation collapse back home, i.e. it is the optimal
migration rate (see Fig. \ref{fig:howto_find_threshold}(middle)

Eventually, one can ask how the outbreak of cooperation $(f_{\rm
  thr}^{K},m^{K})$ is affected if instead of the two strategies TFT,
ALL-D three or four strategies are considered. In general, the presence
of both ALL-C and A-TFT favors the invasion of ALL-D at the end, whereas
the presence of A-TFT favors the invasion of TFT -- thus we can expect
that the thresholds are slightly higher or lower in such cases.  A more
involved discussion is presented in \ref{divers}.

\section{Conclusions}
\label{conclusions}

Before summarizing our findings, we wish to comment on a related strand
of literature about multi-level selection in populations
\citep{Traulsen2006b,Traulsen2008}. There, quite similar to the setup
presented in our paper, a population consisting of several groups
(subpopulations) was considered where interaction only occurs between
agents of the same subpopulation. The fitness of agents was determined by
the payoff obtained from an evolutionary game, simply chosen as a
non-iterative Prisoner's Dilemma, and their reproduction was assumed to
be dependent on their payoff according to a Moran process
\citep{Traulsen2005}. Further a stochastic dynamics was considered. 

In addition to these differences (PD, Moran process, stochastic
dynamics), also a separate group dynamics was assumed. Groups can split
when reaching a certain size, which denotes a amplified replication
process at the group level. Precisely, groups that contain fitter agents
reach the critical size faster and, therefore, split more often. This
concept leads to selection among groups, although only individuals
reproduce. It allows for the emergence of higher level selection (group)
from lower level reproduction (agents).  It was shown
\citep{Traulsen2006b,Traulsen2008} that cooperation in such a setting is
favored if the size of groups is small whereas the number of groups is
large. Migration does \emph{not} support the outbreak of cooperation in
this model because it favors the invasion of defectors rather than of
cooperators.

In contrast to these investigations, we have contributed to the analysis
of evolutionary (deterministic) IPD games in two different domains:
\begin{itemize}
\item[(i)] For the meanfield limit, represented by a panmictic population
  where each agent interacts with $N-1$ other agents $n_{g}$ times, a
  quantitative investigation of the basins of attraction was
  provided. Dependent on the initial mix of two cooperative (ALL-C, TFT)
  and two defective (ALL-D, A-TFT) strategies and the number of
  interactions $n_{g}$, we could derive the critical conditions
  (threshold frequencies) under which the outbreak of cooperation can be
  expected in a panmictic population. A detailed analysis of the two,
  three and four strategy cases helped to better understand the
  contribution of each strategy on the final outcome.
\item[(ii)] Using the ``unpertubed'' (or meanfield) limit as a reference
  point, the impact of migration on the outbreak of cooperation in a
  spatially distributed system could be quantified. We found that there
  is a bandwidth of optimal migration rates which lead to the induction
  (or ``export'') of cooperation to a number of islands which otherwise
  would converge to defection. We were able to determine the critical
  conditions $f_{\rm thr}^{K},m^{K}$ which guarantee the outbreak of
  cooperation in a worst case scenario. Remarkably, the relative effort
  $f_{\rm thr}/K$ to export cooperation to defecting islands was
  decreasing with $K$ in some range and always stayed below the critical
  value for an isolated island. I.e. effectively it became easier to turn
  defecting into cooperating islands, provided the optimal migration rate
  was reached.
\end{itemize}
It is important to note that the outbreak of cooperation is not enforced
by a maximal migration rate but, as stated above, by a small range of
optimal migration rates. This is because migration, different from other
approaches, is not seen as an unidirectional dynamics, where agents just
move to a ``better'' place. Instead, our model is based on the assumption
that it is a \emph{bidirectional} process, which we deem a more realistic
assumptions when considering a dynamics over many generations. In fact,
agents, or their offsprings, which have obtained certain skills or wealth
at their immigration country, quite often decide to start new buisness
back home at their origin country -- because their new capabilities give
them a considerable advantage there, while it is only a marginal
advantage in their immigration country. This assumption is in line with
many policies about immigration/education of foreigners in a country, to
give indirect support for development and to allow future business with
those countries, where some agents with positive experience have
resettled. At the end, as we have shown in our paper, it remarkably helps
to spread ``cooperation'' globally, if the number of breeding places for
such a strategy is increased (above a minimum threshold).

\subsection*{Acknowledgement}

This work on the outbreak of cooperation was supported by the COST-Action
MP0801 \emph{Physics of Competition and Conflicts}. F.S.  announces
financial support from the Swiss State Secretariat for Education and
Research SER C09.0055.

\appendix

\section{Calculating the average payoff}
\label{appA}

In order to illustrate \eqn{strat-payoff}, let us look at two agents $i$
and $j$. If both of them play strategy $z^{(1)}$, TFT, then, during the
$n_{g}$ different ``iterations'' , their choices are as follows:
\begin{equation}
  \label{1-1}
  \begin{array}[c]{ccccccccccccccccc}
    & & 1 & 2 & 3 & 4 &\cdots & n_{g}\\
    & \mathrm{payoff}\;\, a_{i}^{(11)}: & R & R & R & R & \cdots & R \\
    i & (TFT): & C & C & C & C& \cdots & C\\
    j & (TFT): & C & C & C & C& \cdots & C \\
    & \mathrm{payoff}\;\, a_{j}^{(11)}: & R & R & R & R & \cdots & R
  \end{array}
\end{equation}
Consequently, the average payoff for both agents $i$ and $j$ is
$a_{i}^{(11)}=a_{j}^{(11)}=R$. If agent $i$ playing TFT meets with an
agent $j$ playing strategy $z^{(2)}$ (ALL-D) the series of choices would
be:
\begin{equation}
  \label{1-2}
  \begin{array}[c]{rrcccccccccccccccc}
    & &  1 & 2 & 3 & 4&  \cdots & n_{g}\\
    & \mathrm{payoff}\;\, a_{i}^{(12)}: & S & P & P & P & \cdots & P \\
    i & (TFT): & C & D & D &  D& \cdots & D \\
    j & (ALL-D): & D & D & D &  D& \cdots & D \\
    & \mathrm{payoff}\;\, a_{j}^{(21)}: & T & P & P & P & \cdots & P 
  \end{array}
\end{equation}
which leads to an average payoff for agent $i$ of $a_{i}^{(12)}=[S +
(n_{g}-1)P]/n_{g}$, while the average payoff for agent $j$ is
$a_{j}^{(21)}=[T +(n_{g}-1)P]/n_{g}$.

If agent $i$ playing A-TFT interacts with agent $j$ playing ALL-D, the
series of choices reads:
\begin{equation}
  \label{4-2}
  \begin{array}[c]{rrcccccc}
    &  &  1 & 2 & 3 & 4 &  \cdots & n_{g} \\
    & \mathrm{payoff}\;\, a_{i}^{(42)}: & P & S & S & S & \cdots & S \\
    i & (A-TFT): & D & C & C &  C& \cdots & C \\
    j & (ALL-D): & D & D & D &  D & \cdots & D  \\
    & \mathrm{payoff}\;\, a_{j}^{(24)}: & P & T & T & T & \cdots & T   
  \end{array}
\end{equation}
That means the average payoff for agent $i$ is $a_{i}^{(42)}=[P +
(n_{g}-1)S]/n_{g}$, while the average payoff for agent $j$ is
$a_{j}^{(24)}=[P + (n_{g}-1)T]/n_{g}$.  The remaining cases can be
explained similarly. Only the three more complex cases are given
below. $a_{i}^{(14)}$ follows from:
\begin{equation}
  \label{1-4}
  \begin{array}[c]{rrcccccccccccccccc}
    &  &  1 & 2 & 3 & 4 & 5 & 6 & 7 & 8 & \cdots \\
    & \mathrm{payoff}\;\, a_{i}^{(14)}: & S & P & T & R & S & P & T & R & \cdots\\
    i & (TFT): & C & D & D &  C& C & D & D& C & \cdots  \\
    j & (A-TFT): & D & D & C &  C & D & D & C & C & \cdots  \\
    & \mathrm{payoff}\;\, a_{j}^{(41)}: & T & P & S & R & T & P & S & R & \cdots
  \end{array}
\end{equation}
Here, the average payoff of agent $i$ results from a repeated payoff
series $S\,P\,T\,R$, and the value of $n_{g}$ defines when this series is
truncated. Similarly, the series for $a_{i}^{(41)}$ is $T\,P\,S\,R$.
Thus the first term of $a_{i}^{(rq)}$, \eqn{a-qr} calculates all
\emph{completed} payoff series ($n_{g}\,\mathrm{div}\,4$), while the
second term calculates the remaining payoffs ($n_{g}\,\mathrm{rem}\,4$).

For $a_{i}^{(44)}$, we find eventually:
\begin{equation}
  \label{4-4}
  \begin{array}[c]{rrcccccccccccccccc}
    &  & 1 & 2 & 3 & 4 & \cdots \\
    & \mathrm{payoff}\;\, a_{i}^{(44)}: & P & R & P & R & \cdots \\
    i & (A-TFT): & D & C & D &  C & \cdots  \\
    j & (A-TFT): & D & C & D &  C & \cdots  \\
    & \mathrm{payoff}\;\, a_{j}^{(44)}: & P & R & P & R & \cdots
  \end{array}
\end{equation}
Here, the period of the payoff series is just 2 instead of 4, and the
expression for $a_{i}^{(44)}=a_{j}^{(44)}$, \eqn{a-qr} follows
accordingly.

\section{Two strategies: TFT and ALL-D}
\label{2d}
We assume that the population consists of only two strategies:
$z^{(1)}=$TFT and $z^{(2)}=$ALL-D, \eqn{strat}. This is an interesting
combination since defection is known to be the only equilibrium in an
one-shot PD game, while TFT fares well in a repeated PD interaction,
given there is a critical number of encounters (discussed below).

The average payoff $a_{i}^{(rq)}$ received by agent $i$ playing strategy
$z^{(q)}$ ($q \in 1,2$) results from \eqn{strat-payoff}. Choosing
$n_{g}=4$ we find:
\begin{equation}
  \label{matrix-2}
  \left[\begin{array}{cc}
      a^{(11)} & a^{(12)} \\
      a^{(21)} & a^{(22)}
    \end{array}\right]
  = 
  \left[\begin{array}{cc}
      3.0 & 0.75 \\
      2.0 & 1.0
    \end{array}\right]
\end{equation}
Applying \eqs{a-bar}{a-per}, we find for the total average payoff:
\begin{equation}
  \label{a-tot-2}
  \bar{a}=a^{(11)} {f^{(1)}}^2 +(a^{(12)}+a^{(21)}) f^{(1)} f^{(2)}+a^{(22)} {f^{(2)}}^2 
\end{equation}
On the other hand, it follows from the stationary condition (\ref{stat}):
\begin{equation}
  \label{eq:simul}
  \begin{aligned}
    \bar{a}f^{(i)} - \Big(a^{(i1)} f^{(1)} + a^{(i2)} f^{(2)}\Big)
    f^{(i)}
    &=0 \quad  (i=1,2) \\
    f^{(1)}+f^{(2)}&=1
  \end{aligned}
\end{equation}
The combined \eqs{a-tot-2}{eq:simul} have to be solved simultaneously for
the possible stationary frequencies $f^{(1)}$, $f^{(2)}$.  As the result
we find:
\begin{equation}
  \label{separat-2}
  \begin{array}{cl}
    (i)\quad & f^{(1)}=0,\; f^{(2)}=1  \quad \mbox{(stable)} \\
    & 
    f^{(1)}=1,\;  f^{(2)}=0 \quad \mbox{(stable)}\\
    (ii) \quad & f^{(1)}_{\mathrm{thr}}=0.2, \; f^{(2)}_{\mathrm{thr}}=0.8 
    \quad (\mbox{for} \;\;n_{g}=4)\quad \mbox{(unstable)}
  \end{array}
\end{equation}
Solution (i) implies that either strategy ALL-D or TFT invades the
population completely, which are the two stable attractors of the system
in the case of only two strategies.  On the other hand, solution (ii)
describes the coexistence of the two strategies with different
frequencies within the total population. In the given case, it is is an
unstable point attractor that separates the basins of two stable
attractors, and therefore acts as a separatrix here.  I.e., for an
initial frequency $f^{(1)}(0)\leq 0.2$ of strategy $z^{(1)}=$TFT, the
dynamics of the system will converge into a stationary state that is
entirely dominated by strategy ALL-D (where the payoff per agent is
$\bar{a}=P=1$), whereas in the opposite case the population will entirely
adopt strategy TFT (average payoff $\bar{a}=R=3$).

The threshold value $f^{(s)}_{\mathrm{thr}}$ strongly decreases with the
number of encounters $n_{g}$, as shown in \pic{fc-3} for the case of four
strategies.  This raises the question about the critical $n_{g}$ for
which TFT could invade the whole population. Let us compare the average
payoffs of the strategies $z^{(1)}=$ TFT and $z^{(2)}=$ ALL-D:
\begin{equation}
  \label{compare}
  \begin{aligned}
    a^{(1)}=&a^{(11)}f^{(1)}+a^{(12)}f^{(2)} \\
    a^{(2)}=&a^{(12)}f^{(1)}+a^{(22)}f^{(2)}
  \end{aligned}
\end{equation}
The respective values of $a^{(rq)}$ can be calculated from
\eqn{strat-payoff} for different numbers of $n_{g}$. We find that only
the elements $a^{(12)}$ and $a^{(21)}$ change with $n_{g}$, as follows:
\begin{equation}
  \label{a-1221}
  \begin{array}{r|c|c|c|c }
    n_{g} & 1 & 2 & 3 & \cdots \\ \hline
    a^{(12)} & 0.0 & 0.5 & 0.677 & \cdots \\ 
    a^{(21)} & 5.0 & 3.0 & 2.333 & \cdots
  \end{array}
\end{equation}
With $a^{(11)}=3.0$ and $a^{(22)}=1.0$, we find for $n_g=1$ that
$a^{(21)}>a^{(11)}, a^{(22)}> a^{(12)}$. This implies that
$a^{(2)}>a^{(1)}$ for all possible initial frequencies of the two
strategies in the population. Thus, from the dynamics of \eqn{a-dyn}
(fitness-proportional selection) the extinction of the strategy TFT
results, in agreement with the known results.  For $n_g=2$ we find
$a^{(21)}=a^{(11)}$, $a^{(22)}>a^{(12)}$, which again implies that
$a^{(2)}>a^{(1)}$ for all possible initial frequencies of the two
strategies in the population, i.e. the extinction of strategy TFT.
However, for $n_g=3$, we find $a^{(21)}<a^{(11)}$, $a^{(22)}>a^{(12)}$.
Thus, there exist some initial frequencies of the two strategies for
which $a^{(1)}>a^{(2)}$ results. I.e., we can conclude that for the given
payoff matrix 
$n_{g}^{\mathrm{thr}}=3$ is the \emph{threshold value} for the number of
interactions between each two agents that may lead to the invasion of
cooperation in the whole population.

\section{Three strategies: TFT, ALL-D and ALL-C}
\label{3d}

When strategy $z^{(3)}=$ ALL-C is added to the previous strategies
$z^{(1)}=$ TFT and $z^{(2)}=$ ALL-D, agents playing ALL-D will benefit
from agents playing ALL-C. Therefore, if the frequency of agents playing
ALL-C is increased in the initial population, it can be expected that the
basin of attraction of the TFT strategy will shrink, while the basin of
attraction of the ALL-D strategy will grow, compared to the case of two
strategies, discussed in \ref{2d}.  The stationary frequencies have now
to be calculated from the following set of coupled equations that result
from \eqs{a-bar}{a-per} and the stationary condition (\ref{stat}):
\begin{equation}
  \label{eq:steady_3}
  \begin{aligned}
    \bar{a}=\sum_{s=1}^3 a^{(s)} f^{(s)}=
    \sum_{s=1}^3 \left(\sum_{r=1}^3 a^{(rs)} f^{(r)} \right) f^{(s)} \\
    \bar{a} f^{(i)} - \Big(a^{(i1)}f^{(1)}+a^{(i2)} f^{(2)}
    +a^{(i3)}f^{(3)}\Big)f^{(i)} &=0 \quad (i=1,2,3) \\
    f^{(1)}+f^{(2)}+f^{(3)}&=1
  \end{aligned}
\end{equation}
The matrix elements $a^{(rq)}$ can be calculated from
\eqn{strat-payoff}. With $n_{g}=4$ we find the following stationary
solutions:
\begin{equation}
  \label{separat-3}
  \begin{array}{cl}
    (i)\quad & f^{(1)}=1,\; f^{(2)}=0, \; f^{(3)}=0  \quad \mbox{(stable)}\\ 
    & f^{(1)}=0,\; f^{(2)}=1,\; f^{(3)}=0 \quad \mbox{(stable)}\\
    & f^{(1)}=0,\; f^{(2)}=0,\; f^{(3)}=1 \quad \mbox{(unstable)}\\
    (ii)\quad & f^{(1)}_{\mathrm{thr}}=0.2,\; f^{(2)}_{\mathrm{thr}}=0.8,\;
    f^{(3)}=0 \quad (\mbox{for} \;\;n_{g}=4)\quad \mbox{(unstable)}\\
    & f^{(1)}=1.0-f^{(3)}\; (0\leq
    f^{(3)}\leq 1),\; f^{(2)}=0 \quad \mbox{(stable)}
  \end{array}
\end{equation}
We note that a mean-field analysis of the three-strategy case was also
dicussed in \citep{Vega-Redondo2003} and, assuming extensions such as
mutations and fluctuations, analysed further in \citep{Bladon2010,
  Imhof2005b}.  We consider the deterministc case here. Solutions (i)
imply that either strategy TFT or ALL-D or ALL-C invades the entire
population. We note that only the first two solutions are stable ones,
while the last point-attractor $f^{(3)}=1$, $f^{(1)}=f^{(2)}=0$ is an
unstable one, because any small pertubation (i.e. the invasion of one
defecting agent) will transfer the cooperating system into a defecting
one.

The first of the solutions (ii) describing coexisting strategies is
already known from the investigation in \ref{2d} to be an unstable one.
In the absense of the third strategy, it defines the separatrix point.
The second solution (ii) however is a stable one, indicating that both
agents playing TFT and ALL-C can coexist in the panmictic population.
Note that there is neither a stable nor an unstable coexistence of all
three strategies.

The separatix that divides the different basins of attraction is now a
line in a two-dimensional space of the initial frequencies.  But
different from \ref{2d} the stationary solutions of \eqn{separat-3} do
not give further information about the separatices. In order to calculate
the different basins of attraction, we therefore have numerically solved
\eqn{a-dyn} for the full range of initial frequencies: $0 \leq f^{(1)}(0)
\leq 1$, $0 \leq f^{(2)}(0) \leq 1$, $f^{(3)}(0) = 1 - f^{(1)}(0) -
f^{(2)}(0)$, and have evaluated the average payoff in the stationary
limit. If $\bar{a}=P$, obviously the whole population has adopted
strategy $z^{(2)}=$ ALL-D.  Similarly, if $\bar{a}=R$ and $f^{(1)}=1$,
the whole population has adopted strategy $z^{(1)}=$ TFT. However, if
$\bar{a}=R$ and $f^{(1)}<1$, then there is a coexistence of agents
playing strategy $z^{(1)}=$ TFT and $z^{(3)}=$ ALL-C.
 
The basins of attraction are shown in \pic{fig:basin_3} for two different
values of $n_g=4$ and $n_g \to \infty$. In the latter case,
$a^{(12)}=a^{(21)}=1$ results for the payoffs in \eqn{strat-payoff}.  For
$n_{g}=4$ we can distinguish between three different regions in
\pic{fig:basin_3}. Region $A$ denotes the range of initial frequencies
$f^{(s)}(0)$ that lead to the adoption of the $z^{(2)}=$ ALL-D strategy
in the whole population, region $B$ denotes the range of initial
frequencies that lead to the adoption of the $z^{(1)}=$ TFT strategy in
the whole population, while region $C$ denotes the range of initial
frequencies that lead to the coexistence of both $z^{(1)}=$ TFT and
$z^{(3)}=$ ALL-C strategies.
\begin{figure}[htbp]
  \centerline{
    \includegraphics[width=6.0cm]{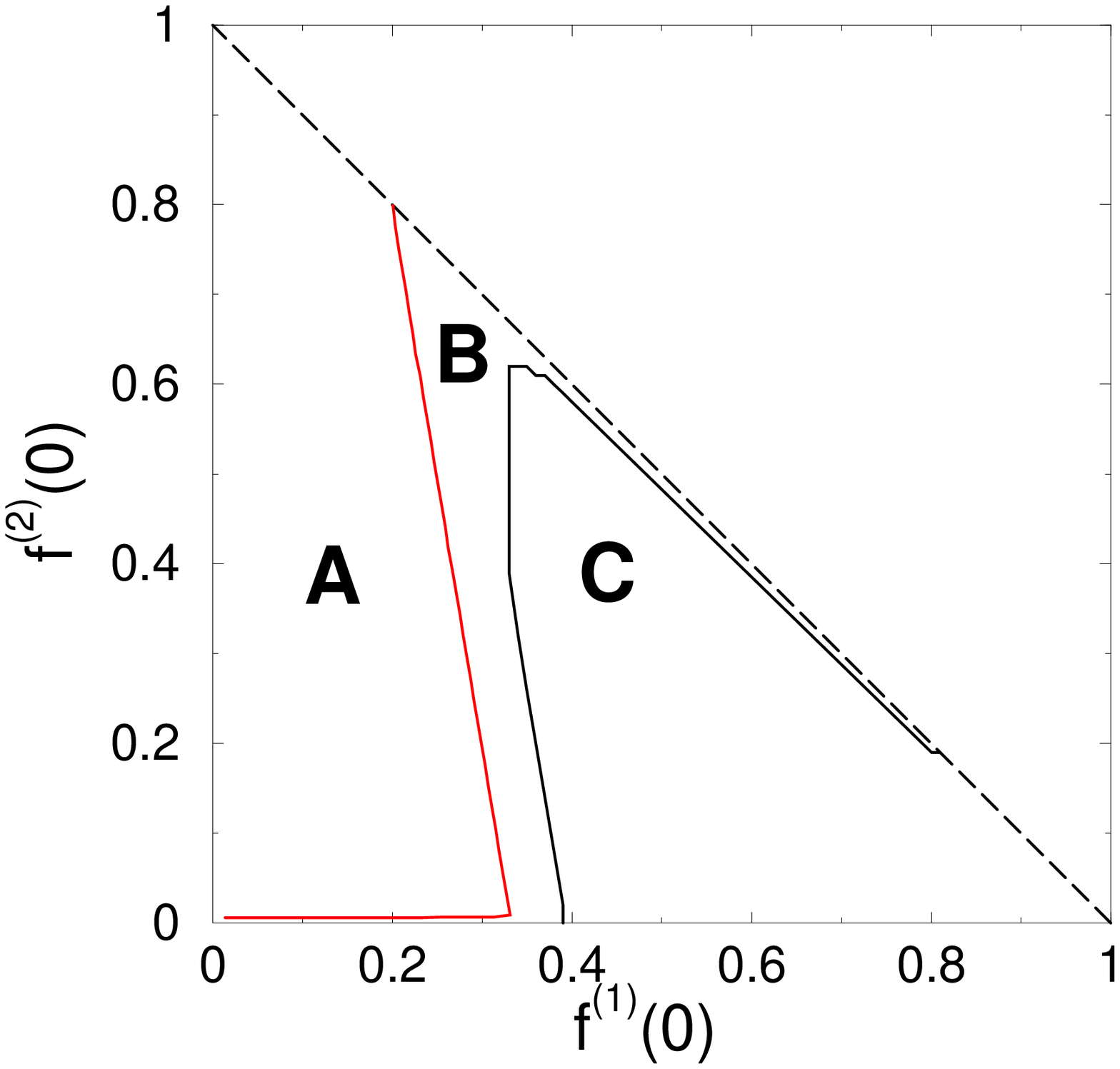} 
\hspace{1cm}
    \includegraphics[width=6.0cm]{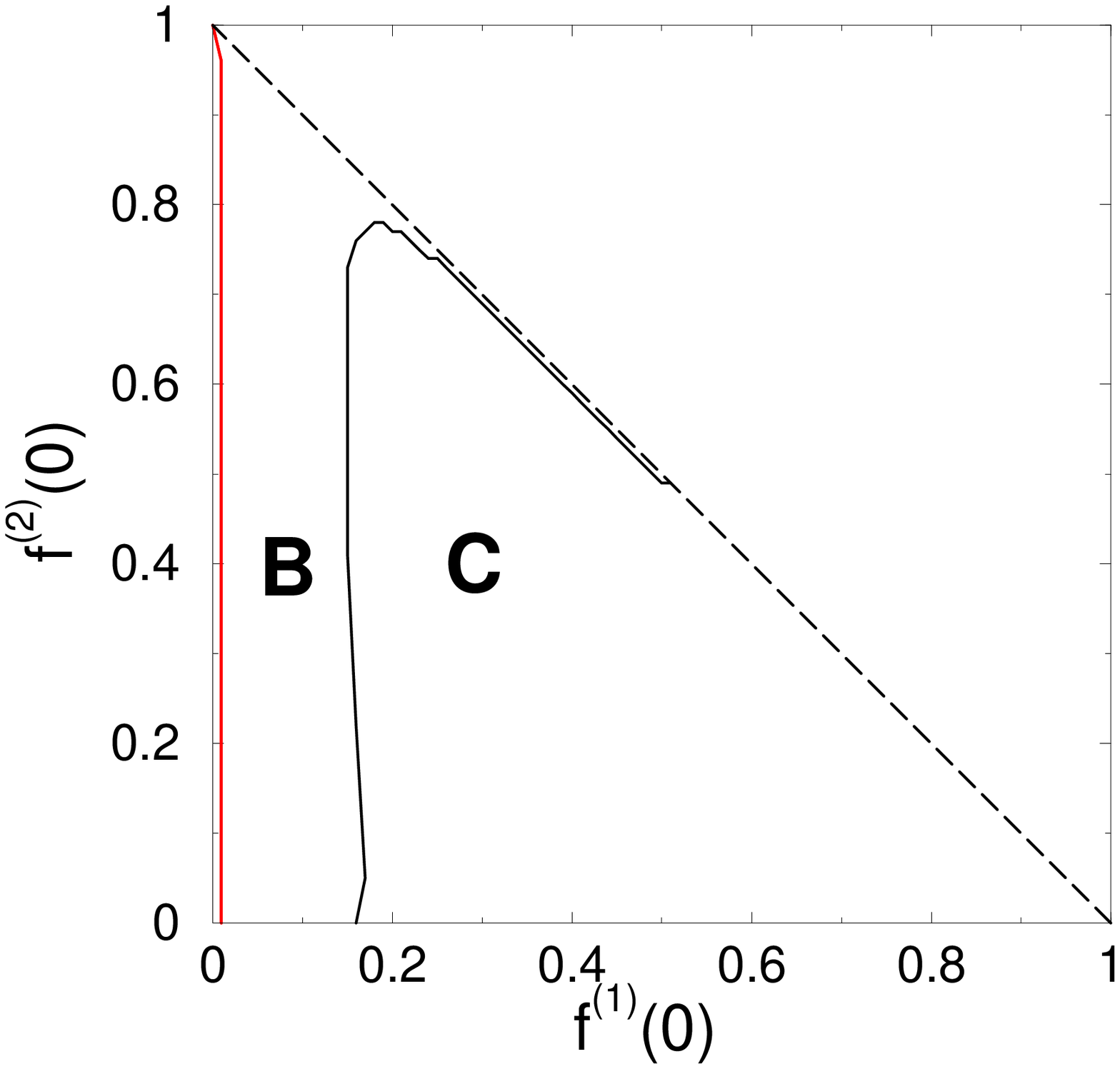}
  }
    \caption{Basins of attraction, i.e. range of initial frequencies
      $f^{(s)}(0)$ that lead to a particluar stable solution,
      \eqn{separat-3}. 
$A$:  adoption of $z^{(2)}=$ ALL-D strategy
in the whole population, $B$:  adoption of $z^{(1)}=$ TFT strategy
in the whole population,  $C$: coexistence of both $z^{(1)}=$ TFT and
$z^{(3)}=$ ALL-C strategies. $f^{(3)}(0)=1-f^{(1)}(0)-f^{(2)}(0)$. 
(left)  $n_g=4$, (right) $n_g \to\infty$. 
\label{fig:basin_3}}
\end{figure}

Since regions $B$ and $C$ both describe the adoption of cooperating
strategies in the population, the most interesting line in
\pic{fig:basin_3} is the separatrix between region $A$ (all defectors)
and region $B$ (all cooperators).  We can easily interpret this line
based on our previous analysis of the two-strategy case, \ref{2d}.  The
diagonal in \pic{fig:basin_3} represents $f^{(1)}+f^{(2)}=1$, i.e. a
population with only two strategies, $z^{(1)}=$ TFT and $z^{(2)}=$ ALL-D.
Thus the separatrix line between regions $A$ and $B$ starts from the
separatrix point $ f^{(1)}=0.2$, $f^{(2)}=0.8$, $f^{(3)}=0$,
\eqn{separat-3}. Further below the diagonal, the frequency $f^{(3)}$ of
the strategy $z^{(3)}=$ ALL-C increases in the initial population, which
in turn increases the threshold frequency $f^{(1)}$ necessary for the
invasion of the $z^{(1)}=$ TFT strategy. In a certain range of
frequencies, the separatrix line between defection and cooperation can be
described by a linear relation, found numerically as:
\begin{equation}
  \label{line}
  f^{(2)}=-6.1538f^{(1)}+2.0308 \quad \mbox{for}\;\;
  0.2 \leq f^{(1)} \leq 0.31   
\end{equation}
However, we notice that the separatrix line between regions $A$ and $B$
never hits the x-axis. For very low values of $f^{(2)}(0)$, i.e. close to
the x-axis, it makes a sharp turn towards the origin. This means that for
a vanishing initial frequency of ALL-D there will be no route to the
respective attraction region $A$, which is obviously correct.

The influence of the parameter $n_{g}$ is shown by comparing the left
($n_{g}=4$) and the right part ($n_{g}=\infty$) of \pic{fig:basin_3}. In
the latter case the basin of attraction $A$ (exclusive domination of strategy
$z^{(2)}=$ ALL-D) becomes very small.
In order to further quantify the influence of $n_{g}$ on the dominating
strategies in the stationary limit, we have also calculated the relative
size of each basin of attraction. If $F_A$, $F_B$ and $F_{C}$ denote the area
of the regions $A$, $B$ and $C$ in \pic{fig:basin_3}, the relative sizes
are defined as follows:
  \begin{equation}
    \label{area}
    a=\frac{F_A}{F_{tot}}; \;b=\frac{F_B}{F_{tot}}; \;c=\frac{F_{C}}{F_{tot}}; 
    \;d=\frac{F_B+F_{C}}{F_{tot}};
    \quad 
    F_{tot}=F_A+F_B+F_{C}
  \end{equation}
  The results are shown in \pic{fig:bar_3} for the two different values
  of $n_{g}$.
  \begin{figure}[htbp]
    \centerline{\includegraphics[width=6.0cm]{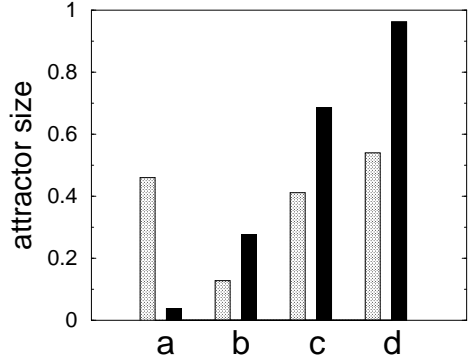}}
\caption{Relative size, \eqn{area} of the basins of attraction
  shown in \pic{fig:basin_3}. The left bars (shaded area) refer to
  $n_{g}=4$ (\pic{fig:basin_3} left), while the right bars (black area)
  refer to $n_{g}\to \infty$ (\pic{fig:basin_3} right). Thus, the change
  indicates the influence of $n_{g}$ on the size of the basins of attraction.
  \label{fig:bar_3}}
\end{figure}

In \pic{fig:bar_3}, $d$, \eqn{area} denotes the relative size of the
basin of attraction for cooperation resulting from both solutions,
domination of $z^{(1)}=$ TFT and coexistence of $z^{(1)}=$ TFT and
$z^{(3)}=$ ALL-C. As we see, for $n_g=4$, the cooperative basin $d$ only
has about the same size as the basin of attraction for defection $a$, that
mean that about half of the possible initial conditions will lead to a
population of defecting agents, at the end. Only for $n_g \to \infty$ the
size of the defection basin becomes insignificant as compared to the
cooperative basin. This again explains the role of $n_g$ in influencing
cooperation.

\section{Threshold frequencies for three and four strategies}
\label{divers}

\begin{figure}[htbp]
  \centerline{\includegraphics[width=6.0cm]{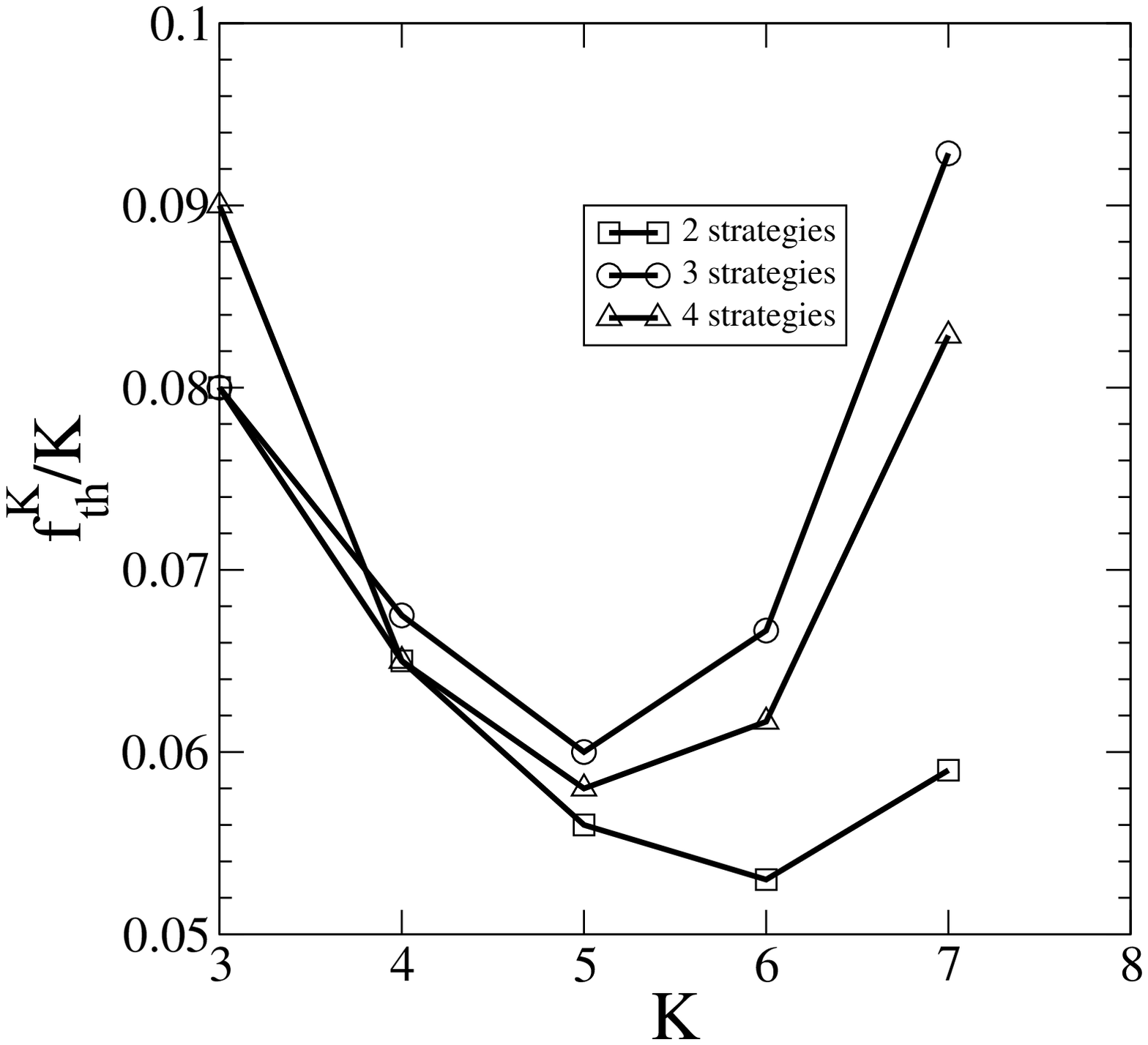}
    \hspace{1cm}
    \includegraphics[width=6.0cm]{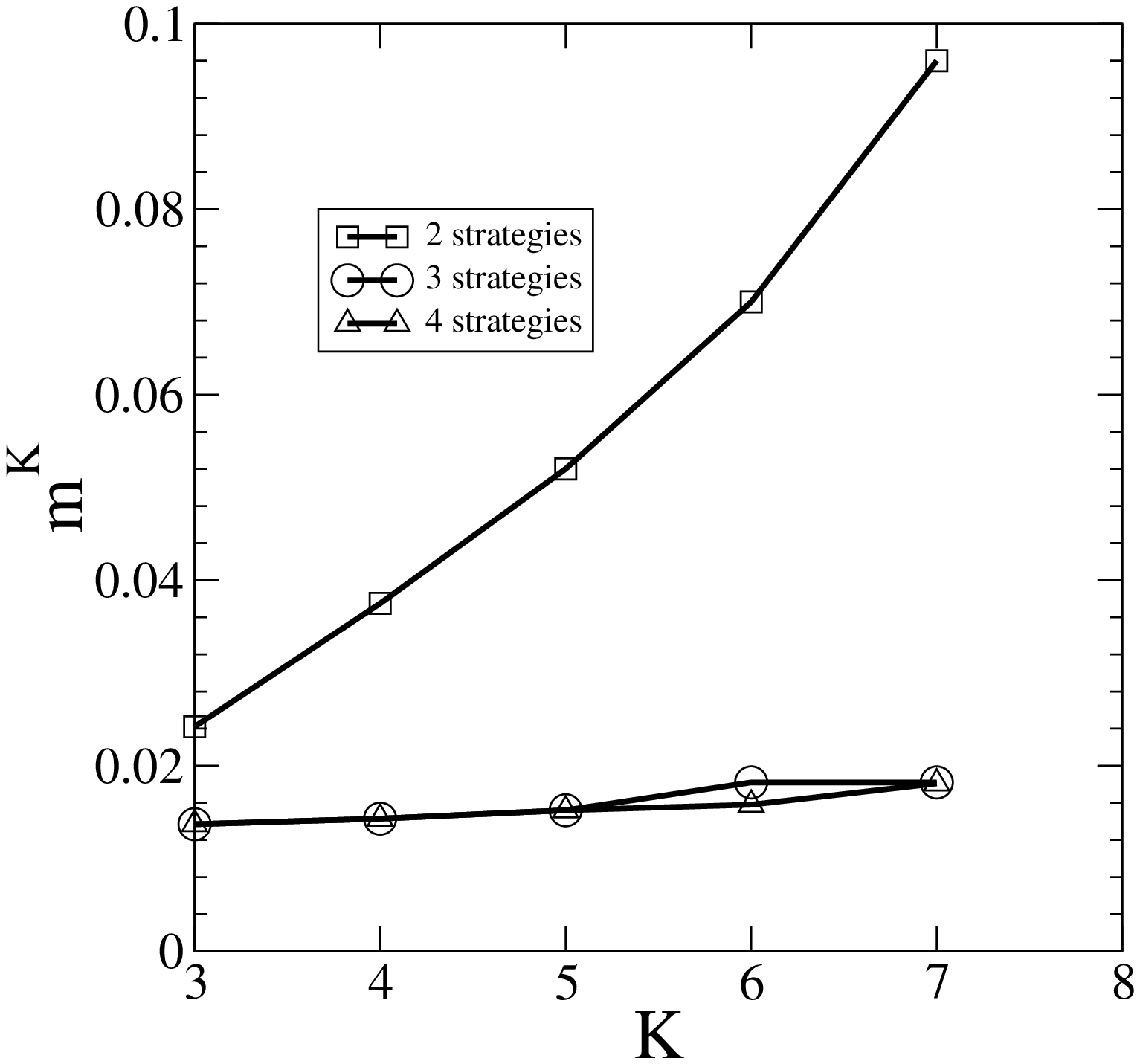}}
  \caption{(left) Threshold $f_{\rm thr}^K$ versus number of islands
    $K$. (right) Migration rate $m$ versus number of islands
    $K$. Migration can only promote cooperation if the number of islands
    is $K\leq 7$}
  \label{fig:thres_and_migr_vs_K_all}
\end{figure}

In Section \ref{sec:threshold}, we computed the critical conditions for
the outbreak of cooperation $(f_{\rm thr}^K, m^K)$ in the presence of only
two strategies, ALL-D and TFT. If we consider the two additional
strategies ALL-C and A-TFT, eq. (\ref{strat}), these critical conditions
change dependent on the initial values of the four strategies and their
distribution on the different islands. Thus, instead of a complete
investigations we only discuss the following sample configurations with
$\sum _{s}f_{k}^{(s)}(0)=1$: 
\begin{itemize}
\item[(a)] \emph{three} strategies, TFT, ALL-D, ALL-C. $f_{1}^{(1)}(0)>0,
  f_{1}^{(2)}(0)=1-f_{1}^{(1)}(0)$, $f_{2}^{(3)}(0)=1$,
  $f_{k}^{(2)}(0)=1$ with $k\geq 3$.
\item[(b)] \emph{four} strategies, TFT, ALL-D, ALL-C,
  A-TFT. $f_{1}^{(1)}(0)>0, f_{1}^{(2)}(0)=1-f_{1}^{(1)}(0)$,
  $f_{2}^{(3)}(0)=1$, $f_{3}^{(4)}=1$, $f_{k}^{(2)}(0)=1$ with $k\geq 4$.
\end{itemize}
These initial conditions imply that $K\geq 3$. The results of extensive
calculations of the relative effort $f^{K}_{\rm thr}/K$ and the critical
migration rate $m^{K}$ are shown in
Fig. \ref{fig:thres_and_migr_vs_K_all} and can be compared to the two
strategy case, eq.
(\ref{eq:threshold}). 

Again, we notice that the relative effort to invade other islands by
agents playing TFT is non-monotonously dependent on $K$ and always stays
below the threshold values observed without migration, as given in
Sect. \ref{3.1}, \ref{3d}. Regarding the impact of the different
strategies on the outbreak of cooperation, we see that in the presence of
ALL-C ALL-D benefits more than TFT in terms of payoff 
which raises the threshold frequency.  However, adding A-TFT benefits
ALL-D less than adding ALL-C, which lowers the threshold frequency.  This
explains why the curve for the \emph{four} strategy case is in between
the curves corresponding to \emph{two} and \emph{three} strategies.
From \pic{fig:thres_and_migr_vs_K_all}(right) the optimal migration rate
is found to be almost constant for three and four strategies in contrast
to the two strategy case.


\begin{thebibliography}{10}
\providecommand{\urlprefix}{}
\expandafter\ifx\csname urlstyle\endcsname\relax
  \providecommand{\doi}[1]{doi:\discretionary{}{}{}#1}\else
  \providecommand{\doi}{doi:\discretionary{}{}{}\begingroup
  \urlstyle{rm}\Url}\fi

\bibitem{Axelrod:81}
Axelrod, R. and Hamilton, W., The evolution of cooperation, \emph{Science}
  \textbf{211} (1981) 1390--1396.

\bibitem{Bladon2010}
Bladon, A.~J., Galla, T., and McKane, A.~J., {Evolutionary dynamics, intrinsic
  noise and cycles of co-operation}, \emph{Physcal Review E} \textbf{81} (2010)
  066122--(1--12).

\bibitem{Cohen:99}
Cohen, M.~D., Riolo, R.~L., and Axelrod, R., The emergence of social
  organization in the prisoner's dilemma: How context-preservation and other
  factors promote cooperation, Technical Report 99-01-002, Santa Fe Institute
  (1999), working paper.

\bibitem{Doebeli:97}
Doebeli, M., Blarer, A., and Ackermann, M., Population dynamics, demographic
  stochasticity, and the evolution of cooperation, \emph{Proc. Natl Acad. Sci,
  USA} \textbf{94} (1997) 5167--5171.

\bibitem{Fogel:95}
Fogel, D.~B., On the relationship between the duration of an encounter and
  evolution of cooperation in the iterated prisoner's dilemma,
  \emph{Evolutionary Computation} \textbf{3} (1995) 349--363.

\bibitem{helbing2008migration}
Helbing, D. and Yu, W., Migration as a mechanism to promote cooperation,
  \emph{Adv. Complex Syst} \textbf{11} (2008) 641--652.

\bibitem{helbing2009outbreak}
Helbing, D. and Yu, W., The outbreak of cooperation among success-driven
  individuals under noisy conditions, \emph{Proceedings of the National Academy
  of Sciences} \textbf{106} (2009) 3680.

\bibitem{Hof:88}
Hofbauer, J. and Sigmund, K., \emph{The Theory of Evolution and Dynamical
  Systems} (Cambridge University Press, 1988).

\bibitem{Imhof2005b}
Imhof, L.~A., Fudenberg, D., and Nowak, M.~A., {Evolutionary cycles of
  cooperation and defection.}, \emph{Proceedings of the National Academy of
  Sciences of the United States of America} \textbf{102} (2005) 10797--800.

\bibitem{jiang2010role}
Jiang, L., Wang, W., Lai, Y., and Wang, B., Role of adaptive migration in
  promoting cooperation in spatial games, \emph{Physical Review E} \textbf{81}
  (2010) 036108.

\bibitem{lozano2008mesoscopic}
Lozano, S., Arenas, A., and S{\'a}nchez, A., Mesoscopic structure conditions
  the emergence of cooperation on social networks, \emph{PLoS One} \textbf{3}
  (2008) e1892.

\bibitem{Michael:96}
Michael, M., Natural selection and social learning in prisoner's dilemma,
  \emph{Sociological Methods and Research} \textbf{25} (1996) 103--138.

\bibitem{Nowak:92}
Nowak, M.~A. and Sigmund, K., Tit for tat in heterogeneous populations,
  \emph{Nature} \textbf{355} (1992) 250--253.

\bibitem{Rapoport:96}
Rapoport, A., Prisoner's dilemma: reflections and recollections,
  \emph{Simulation and Gaming} \textbf{26} (1996) 489--503.

\bibitem{roca2009effect}
Roca, C., Cuesta, J., and S{\'a}nchez, A., Effect of spatial structure on the
  evolution of cooperation, \emph{Physical Review E} \textbf{80} (2009) 046106.

\bibitem{schweitzer02:_evolut}
Schweitzer, F., Behera, L., and M{\"u}hlenbein, H., Evolution of cooperation in
  a spatial prisoner's dilemma, \emph{Advances in Complex Systems} \textbf{5}
  (2002) 269--300.

\bibitem{Schweitzer2005}
Schweitzer, F., Mach, R., and M, H., {Agents with heterogeneous strategies
  interacting in a spatial IPD Agent's Strategies}, in \emph{Lecture Notes in
  Economics and Mathematical Systems}, eds. Lux, T., Reitz, S., and Samanidou,
  E., i (Springer, 2005), pp. 87--102.

\bibitem{szabo00:_spatial}
Szabo, G., Antal, T., Szabo, P., and Droz, M., Spatial evolutionary prisoner's
  dilemma game with three strategies and external constraints, \emph{Physical
  Rev. E} \textbf{62} (2000) 1095--1103.

\bibitem{Szabo:98}
Szabo, G. and Toke, C., Evolutionary prisoner's dilemma game on a square
  lattice, \emph{Physical Review E} \textbf{58} (1998) 69--73.

\bibitem{Traulsen2006b}
Traulsen, A. and Nowak, M.~A., {Evolution of cooperation by multilevel
  selection}, \emph{Proceedings of the National Academy of Sciences of the
  United States of America} \textbf{103} (2006) 10952--5.

\bibitem{Traulsen2005}
Traulsen, A., Sengupta, A.~M., and Nowak, M.~A., {Stochastic evolutionary
  dynamics on two levels}, \emph{Journal of theoretical biology} \textbf{235}
  (2005) 393--401.

\bibitem{Traulsen2008}
Traulsen, A., Shoresh, N., and Nowak, M.~A., {Analytical results for individual
  and group selection of any intensity.}, \emph{Bulletin of mathematical
  biology} \textbf{70} (2008) 1410--24.

\bibitem{Vainstein:02}
Vainstein, M.~H. and Arenzon, J.~J., Disordered environments in spatial games,
  \emph{Physical Review E}  (2002).

\bibitem{Vega-Redondo2003}
Vega-Redondo, F., \emph{{Economics and the Theory of Games}}, 1st edn.
  (Cambridge University Press, New York, 2003).

\end{thebibliography}
\end{document}